%% file: ms_21cmfish.tex
\documentclass[fleqn,usenatbib]{mnras}

\usepackage{newtxtext,newtxmath}
\AtBeginDocument{\mathcode`v=\varv} 

\usepackage[T1]{fontenc}
\usepackage{ae,aecompl}

\usepackage{graphicx}	
\usepackage{xspace}	    

\usepackage{caption}    
\usepackage{subcaption}
\usepackage{multirow}

\include{macros}

\graphicspath{{./}{figures/}}
\defcitealias{PlanckCollaboration2018}{P18} 	  	

\title[21-cm Fisher-matrix analysis]{21cmfish: Fisher-matrix framework for fast parameter forecasts from the cosmic 21-cm signal}

\author[C. A. Mason et al.]{Charlotte A. Mason$^{1,2}$\thanks{E-mail: charlotte.mason@nbi.ku.dk},
Julian B. Mu\~noz$^{3,4}$, 
Bradley Greig$^{5,6}$,
Andrei Mesinger$^{7}$,
\newauthor{and Jaehong Park$^{8}$}
\\
$^{1}$Cosmic Dawn Center (DAWN)\\
$^{2}$Niels Bohr Institute, University of Copenhagen, Jagtvej 128, 2200 København N, Denmark\\
$^{3}$Center for Astrophysics $|$ Harvard \& Smithsonian, 60 Garden St., Cambridge, MA 02138, USA\\
$^{4}$The University of Texas at Austin, Department of Astronomy, Austin, TX 78712\\
$^{5}$School of Physics, University of Melbourne, Parkville, VIC 3010, Australia\\
$^{6}$ARC Centre of Excellence for All Sky Astrophysics in 3 Dimensions (ASTRO 3D)\\
$^{7}$Scuola Normale Superiore, 56126 Pisa, PI, Italy\\
$^{8}$School of Physics, Korea Institute for Advanced Study (KIAS), 85 Hoegiro, Dongdaemun-gu, Seoul 02455, Republic of Korea
}

\date{Accepted XXX. Received YYY; in original form ZZZ}

\pubyear{2023}

\begin{document}
\label{firstpage}
\pagerange{\pageref{firstpage}--\pageref{lastpage}}
\maketitle

\begin{abstract}
%
The 21-cm signal from neutral hydrogen in the early universe will provide unprecedented information about the first stars and galaxies. Extracting this information, however, requires accounting for many unknown astrophysical processes.
Semi-numerical simulations are key for exploring the vast parameter space of said processes.
These simulations use approximate techniques such as excursion-set and perturbation theory to model the 3D evolution of the intergalactic medium, at a fraction of the computational cost of hydrodynamic and/or radiative transfer simulations. However, exploring the enormous parameter space of the first galaxies can still be computationally expensive.
Here we introduce {\tt 21cmfish}, a Fisher-matrix wrapper for the semi-numerical simulation {\tt 21cmFAST}. {\tt 21cmfish} facilitates efficient parameter forecasts, scaling to significantly higher dimensionalities than MCMC approaches, assuming a multi-variate Gaussian posterior. Our method produces comparable parameter uncertainty forecasts to previous MCMC analyses but requires $\sim10^4\times$ fewer simulations. This enables a rapid way to prototype analyses adding new physics and/or additional parameters.
We carry out a forecast for HERA using the largest astrophysical parameter space to-date, with 10 free parameters, spanning both population II and III star formation.
We find X-ray parameters for the first galaxies could be measured to sub-percent precision, and, though they are highly degenerate, the stellar-to-halo mass relation and ionizing photon escape fraction for population II and III galaxies can be constrained to $\sim10\%$ precision (logarithmic quantities). Using a principal component analysis we find HERA is most sensitive to the product of the ionizing escape fraction and the stellar-to-halo mass fraction for population II galaxies.
\end{abstract}




\section{Introduction}
\label{sec:intro}

The cosmic 21-cm signal will soon open a new window on the early universe. The signal is a net absorption or emission of 21-cm photons relative to the cosmic microwave background (CMB) by neutral hydrogen gas in the intergalactic medium (IGM) that depends sensitively on the formation redshift and properties of the universe's first stars and galaxies, see e.g., reviews by \citet{Furlanetto2006a,Morales2010,Pritchard2012,Mesinger2019}. When the first stars form ($z\sim30$), their ultra-violet (UV) radiation couples the spin temperature of neutral hydrogen to the cooler kinetic temperature of the gas, via the Wouthuysen-Field effect \citep[e.g.,][]{Hirata2006}, driving a net absorption of CMB photons at 21-cm wavelengths. Later, as the galaxy population builds up, X-ray emission from the first galaxies is expected to heat the IGM, driving the 21-cm signal into emission ($z\simlt15$), and eventually, as hydrogen reionizes, the signal decays to zero ($z\simlt10$). As the formation redshift and abundance of the first stars is highly sensitive to the build up of low mass dark-matter halos ($M_h \sim10^{6-8}\Msun$) and astrophysical feedback effects which govern star formation, the redshift evolution of the 21-cm signal can be used to constrain astrophysical and cosmological properties in the early universe. This will be an exciting way to see beyond the limits of optical and near-infrared telescopes.

In the coming decade, a number of experiments will reach the expected sensitivity required to measure the 21-cm power spectrum at $z>6$, including e.g. the Hydrogen Epoch of Reionization Array \citep[HERA,][]{deBoer2017,HERA2021a} and the Square Kilometer Array \citep[SKA,][]{SKA_Koopmans2015}, while current instruments have set strong upper limits: the Murchison Widefield Array \citep[MWA][]{MWA_Tingay2013,Trott2020} and the Low Frequency Array \citep[LOFAR,][]{vanHaarlem2013,Gehlot2019,Mertens2020}, and a claimed first measurement of the 21-cm global signal has been made at $z\sim18$ (\citealt{Bowman2018a} though c.f. \citealt{Hills2018a,Singh2022}). To interpret these upcoming data, it is imperative to compare observations to models which encompass as much of the unknown physics of the early universe as possible. In recent years there has been a particular effort to create parametric models which describe the key astrophysical and cosmological mechanisms which could affect the 21-cm signal (e.g., {\tt 21cmFAST}\footnote{\url{https://github.com/21cmfast/21cmFAST}}: \citealt{Mesinger2007,Mesinger2011,Murray2020}, {\tt zreion}:~\citealt{Battaglia:2012id}; {\tt ares}: \citealt{Mirocha2014}; {\tt ReionYuga}: \citealt{Majumdar2014,Mondal2017}; {\tt GRIZZLY}: \citealt{Ghara2015}; {\tt SCRIPT}: \citealt{Choudhury2018}; {\tt ASTRAEUS}: \citealt{Hutter2021}; {\tt DRAGONS}: \citealt{Mutch2016}; {\tt SimFAST21}: \citealt{Santos2010}; \citealt{Visbal2010}).

Unknown physics in the early universe determines the formation of the first stars and galaxies and the heating and ionization of the IGM which in turn sets the amplitude and spatial structure of the 21-cm signal. For example: the strength of star formation feedback in low mass halos; Lyman-Werner feedback; the impact of streaming velocities between baryons and dark matter; the typical X-ray emission of early galaxies; and the escape fraction of hydrogen ionizing photons from galaxies all contribute to the strength of the 21-cm signal via their impact on the formation redshifts and typical bias of the first galaxies, the level of inhomogeneous heating of the IGM during cosmic dawn, and the rate at which hydrogen is reionized \citep[e.g.,][]{Munoz2021b}. The degeneracies between some of these effects have been explored \citep[e.g.,][]{Park2019,Qin2020,Qin2021a} but due to computational inefficiencies there has not been a thorough investigation of the degeneracies between effects that govern population II and population III star formation. It is furthermore unclear how much the astrophysical effects will hamper our efforts to understand the underlying cosmology: for example, \citet{Sitwell2014,Jones2021,Munoz2020} have explored the impact of warm, fuzzy, and self-interacting (ETHOS) dark-matter models on the 21-cm signal, respectively, but did not carry out a full exploration of the degeneracies with astrophysical parameters.
Only a few key observables, like the shape of the velocity-induced acoustic oscillations (VAOs, \citealt{Munoz:2019fkt}) are immune to astrophysical uncertainties.

Once 21-cm detections are available, one would ideally map out the multi-dimensional posterior of astrophysics and cosmology using Bayesian inference techniques such as MCMC or Nested sampling \citep[e.g. {\tt 21CMMC}\footnote{\url{https://github.com/21cmfast/21CMMC}}][]{Greig2015,Greig2017b}.  However, mapping out the full posterior is extremely expensive in the high dimensional parameter space of generalized galaxy and cosmological models.  Before having a detection, however, decisions on observational strategies, telescope design, synergies, etc. can be guided by much faster (albeit more approximate) forecasting techniques, such as Fisher-matrix analyses\footnote{Fisher-matrix analyses assume a "fiducial" (i.e. maximum likelihood) parameter set, and computes a multi-variate Gaussian posterior around that point.  These are reasonable approximations, given that we already have a decent estimate of a ``fiducial" model using current data, and previous inference results suggest that the posterior is likely only weakly multi-modal/non-Gaussian (e.g. \citealt{Qin2021a,Qin2021b, HERA2021b}).}.

In this paper, we use a Fisher-matrix analysis to explore an unprecedentedly large {\tt 21cmFAST} astrophysical parameter space. We demonstrate that the Fisher-matrix analysis produces comparable parameter uncertainties to an MCMC with {\tt 21CMMC} (which was restricted to population II galaxy parameters only), recovering the same degeneracies and very comparable error-bars. Having validated our approach we carry out a Fisher-matrix analysis for the largest parameter set to-date in {\tt 21cmFAST}: 10 free parameters, including both population II and population III galaxies over $z \sim 5-30$. Our method reduces the computational cost significantly compared to a typical {\tt 21CMMC} run: requiring a factor $\sim10^4$ fewer individual simulations. We carry out a principal component analysis of our Fisher matrix to determine which combinations of parameters will be most easily constrained by observations. We provide a public release of our code {\tt 21cmfish}\footnote{\url{https://github.com/charlottenosam/21cmfish}}, which is a python wrapper for {\tt 21cmFAST} to perform Fisher-matrix analyses.

This approach will enable us to easily add new parameters, e.g. cosmological and dark-matter parameters, to quickly assess how well upcoming 21-cm experiments will be able to constrain them, which will be the topic of a future work (Verwohlt et al. in prep). The Fisher matrix could also be used for prototyping instrument designs and instrumental noise estimates.

This paper is structured as follows.
Section~\ref{sec:method} describes our method for generating the 21-cm signal with {\tt 21cmFAST}, the astrophysical model and our Fisher-matrix formalism in {\tt 21cmfish}. Section~\ref{sec:results} describes our results of the comparison between the Fisher-matrix analysis and MCMC, and our exploration of the extended model parameter space. We discuss our results in Section~\ref{sec:disc} and summarize our conclusions in Section~\ref{sec:conc}.

In this work we fix the cosmological parameters to the best fit from Planck 2018 data \citep[TT,TE,EE+lowE+lensing+BAO from][]{PlanckCollaboration2018}, and all distances are comoving unless specified otherwise.

\section{Methods} 
\label{sec:method}

Here we provide a brief overview of the astrophysical model (Section~\ref{sec:method_astro}) and the 21-cm signal (Section~\ref{sec:method_21cm}), our Fisher-matrix analysis approach (Section~\ref{sec:method_fisher}), and the generation of simulated observations (Section~\ref{sec:method_21cmsense}). The parametric {\tt 21cmFAST} astrophysical model we use is described by \citet{Park2019,Qin2020,Qin2021a,Qin2021b,Munoz2021b} and we refer readers there for more details.

\subsection{The astrophysical model}
\label{sec:method_astro}

We use the public code {\tt 21cmFAST} \citep{Mesinger2007,Mesinger2011,Murray2020} to model the 21-cm signal. We use version 3.1.0\footnote{\url{https://github.com/21cmfast/21cmFAST/releases/tag/3.1.0}} which includes a parametric model for galaxy formation in both atomic cooling galaxies and molecular cooling galaxies, accounting for all major feedback mechanisms on these galaxies, as well as detailed IGM physics such as inhomogeneous recombinations. The model is fully described by \citet{Park2019,Qin2020} and \citet{Munoz2021b} and we refer the reader there for more details. \citet{Park2019} describes the halo-mass dependent star formation prescription for population II galaxies, \citet{Qin2020} added a prescription for population III galaxies, and \citet{Munoz2021b} self-consistently included feedback effects on population III star formation -- Lyman Werner feedback and relative velocities between dark matter and baryons. Here we provide a brief overview of the astrophysical model.

As halos form hierarchically, we assume that the first generation of stars (population III) form out of pristine gas in molecular cooling galaxies, where gas can cool via transitions in H$_2$ \citep[$T_{\rm vir} \simgt 10^3$\,K, $M_h \simgt 10^6 M_\odot$, e.g.,][]{Barkana2001}, whereas later generations of stars (population II) form in more massive atomic cooling galaxies, where gas cools through atomic transitions ($T_{\rm vir} \simgt 10^4$\,K, corresponding to $M_h \simgt 10^8 M_\odot$), as halos grow and the galaxies' ISM is enriched. {\tt 21cmFAST} models these two galaxy populations separately, as described below. Throughout this work we use PopII or superscript ``II'' to refer to atomic cooling galaxies with population II stars, and PopIII or superscript ``III'' to refer to molecular cooling galaxies with population III stars.

The number densities of these galaxies are given by:
\BE \label{eqn:hmf}
\frac{\mathrm{d}n_g}{\mathrm{d}M_h} = \frac{\mathrm{d}n}{\mathrm{d}M_h} \times f_\mathrm{duty}(M_h)
\EE
where $\frac{\mathrm{d}n}{\mathrm{d}M_h}$ is the halo mass function\footnote{Halo mass functions are created using the Extended Press-Schechter formalism \citep[e.g.,][]{Sheth2001} using a top-hat window function.
}, and $f_\mathrm{duty}$ accounts for the suppression of galaxy formation in halos below a certain mass scale due to a combination of cooling and feedback processes:
\BE \label{eqn:fduty}
f_\mathrm{duty} =
\begin{cases}
    \exp{\left(-\frac{M^\textsc{II}_\mathrm{turn}}{M_h}\right)} & \mathrm{PopII} \\
    \exp{\left(-\frac{M^\textsc{III}_\mathrm{turn}}{M_h}\right)}\exp{\left(-\frac{M_h}{M_\mathrm{atom}}\right)} & \mathrm{PopIII}
\end{cases}
\EE
Here, the second term for population III galaxies produces a smooth transition between the two galaxy populations. The characteristic mass scales where galaxy formation is suppressed, $M^\textsc{II}_\mathrm{turn}, M^\textsc{III}_\mathrm{turn}$, may be set as a free parameter to be inferred \citep[e.g.,][]{Park2019} or modelled physically as a function of redshift and local radiative background strength \citep[e.g.,][]{Munoz2021b}.

For population II atomic cooling galaxies the suppression is determined by atomic cooling and photoionization feedback, and has a characteristic mass scale:
\BE \label{eqn:M_crit_II}
M^\textsc{II}_\mathrm{turn} = \mathrm{max} \left[M_\mathrm{atom}, M_\mathrm{ion} \right]
\EE
where $M_\mathrm{atom}$ is the minimum mass scale of atomic cooling \citep[corresponding to a virial temperature of $\sim10^4$\,K, e.g.][]{Barkana2001,Oh:2001ex}, $M_\mathrm{ion}$ is the mass scale where photoheating feedback from inhomogeneous reionization becomes important \citep[e.g.,][]{Sobacchi2014a}.

For molecular cooling population III galaxies, in addition to the photoionization feedback, we follow \citet{Munoz2021b} (see also \citealt{Fialkov:2012su,Munoz2019}) and include the two dominant sources of feedback that reduce the efficiency of star formation in molecular cooling halos: (i) Lyman-Werner feedback, due to stellar emission at $11.2-13.6$\,eV which effectively photo-dissociates molecular hydrogen, stalling the cooling of gas in population III galaxies and thus impeding star formation; and (ii) dark matter - baryon streaming velocities which reduce the accretion of gas into dark-matter halos, which both slows the formation of halos, due to the lower gravitational potential, and reduces the gas available to form stars. 

The strength of these two effects depends on the Lyman-Werner background flux ($J_{21}$, in units $10^{-21}$\,erg s$^{-1}$ cm$^{-2}$ Hz$^{-1}$ sr$^{-1}$) and the relative velocity between dark matter and baryons ($v_{\rm cb}$ in km s$^{-1}$) as a function of redshift. Therefore,
\BE \label{eqn:M_crit_III}
M^\textsc{III}_\mathrm{turn} = \mathrm{max} \left[M_\mathrm{mol}, M_\mathrm{ion} \right]
\EE
where we use the parameterisation of \citet{Munoz2021b}:
\BE \label{eqn:mturn_iii}
M_\mathrm{mol}(z, v_\mathrm{cb},J_{21}) = \frac{M_\mathrm{mol,crit}}{ (1+z)^{3/2}} f_{v_\mathrm{cb}}(v_\mathrm{cb}) f_\mathrm{LW}(J_{21})
\EE
where $M_\mathrm{mol,crit} = 3.3\times10^7\,\Msun$ is the (no feedback) molecular cooling threshold, corresponding to $T_{\rm vir} \approx 10^3$\,K \citep[e.g.,][]{Barkana2001}, and $f_\mathrm{LW}$ and $f_{v_\mathrm{cb}}$ are the strength of Lyman-Werner and relative-velocity feedback respectively.

The strength of the Lyman-Werner feedback is parameterised as \citep{Machacek2001,Visbal2014}:
\BE \label{eqn:f_LQ}
f_\mathrm{LW}(J_{21}) = 1 + A_\mathrm{LW} \times (J_{21})^{\beta_\mathrm{LW}}
\EE
where $A_\mathrm{LW}$ and $\beta_\mathrm{LW}$ are free parameters. In the following we set $\beta_\mathrm{LW}=0.6$, as simulations predict a small range for this parameter \citep[e.g.,][]{Kulkarni2021,Schauer2021,Skinner_Wise_2020} and leave $A_\mathrm{LW}$ free, and study whether it can be determined from observations. We model the LW intensity, $J_{21}$, in the same way as \citet{Qin2020}, using the PopII- and PopIII-dominated spectral energy distributions (SEDs) from \citet{Barkana2005} to calculate the LW emissivity.

The strength of the relative-velocity feedback is modelled as:
\BE \label{eqn:f_vcb}
f_{v_\mathrm{cb}} (v_\mathrm{cb}) = \left(1 + A_{v_\mathrm{cb}} \frac{v_\mathrm{cb}}{\sigma_\mathrm{cb}} \right)^{\beta_{v_\mathrm{cb}}}
\EE
where $\sigma_\mathrm{cb} \approx 30$\,\kms is the rms velocity~\citep{Tseliakhovich:2010bj}. We follow \citet{Munoz2021b} and use $A_{v_\mathrm{cb}}=1$ and $\beta_{v_\mathrm{cb}}=1.8$, as these values have been recovered consistently in independent simulations \citep{Kulkarni2021,Schauer2021}.

The stellar mass ($M_\star$) to halo mass ratio is described by a power law \citep[e.g.,][]{Moster2013,Behroozi2015a,Sun2016,Mutch2016,Tacchella2018a,Yung2020,Sabti:2021unj,Sabti:2021xvh}:
\BE \label{eqn:mstell_mhalo}
\frac{M_\star}{M_h} = \frac{\Omega_b}{\Omega_m} \times \mathrm{min} \left[ 1,
\begin{cases}
    $\fstarII$ \left(\frac{M_h}{10^{10}M_\odot}\right)^\astarII  & \mathrm{PopII}\\
    $\fstarIII$ \left(\frac{M_h}{10^{7}M_\odot} \right)^\astarIII  & \mathrm{PopIII}\\
\end{cases}
\right]
\EE 
with four free parameters ($\fstarII, \astarII, \fstarIII, \astarIII$) which set the normalization at the pivot mass and low mass slope for galaxies forming population II and population III stars respectively. We note that although the stellar mass -- halo mass relation turns over at high masses \citep[$M_h\sim10^{12}\,M_\odot$, e.g.,][]{Behroozi2013a,Mason2015a}, the radiation fields at these high redshifts are determined by the vast majority of halos which have much lower masses; thus a  single power-law slope suffices for our purposes. This power-law slope encapsulates additional mass-dependent feedback processes, such as supernovae feedback.

We assume the star formation rate in halos is 
related to their dynamical time, which during matter domination scales as $H^{-1}$, to give us
%
\BE \label{eqn:sfr}
\frac{\mathrm{d}M_\star}{\mathrm{d}t} = \frac{M_\star}{\tstar H(z)^{-1}}
\EE 
where \tstar can be a free parameter, but as SFR has the ratio $f_\star/\tstar$ there is a strong degeneracy between these parameters \citep{Park2019}, and \tstar can typically be set to a constant.

Galaxies hosting population II and population III stars can also have different hydrogen ionizing photon escape fractions:
\BE \label{eqn:fesc}
f_\mathrm{esc} = 
\begin{cases}
    $\fescII$ \left(\frac{M_h}{10^{10}M_\odot}\right)^\aesc  & \mathrm{PopII}\\
    $\fescIII$ \left(\frac{M_h}{10^{7}M_\odot} \right)^\aesc  & \mathrm{PopIII}\\
\end{cases}
\EE 
where \fescII and \fescIII set the normalization of the escape fractions and \aesc their power-law slopes. Note that for simplicity, we take \aesc to be the same for each population of galaxies as we assume it is a function of halo mass only.

These early populations of galaxies are also expected to produce X-rays, most likely from high mass X-ray binaries \citep[e.g.][]{Furlanetto2006a,Mineo2011,McQuinn2012,Fragos2013}. We assume the specific X-ray luminosity is proportional to the galaxy SFR and the X-ray SED is a power law with energy spectral index $=-1$, such that the specific X-ray luminosity per unit SFR is:
\BE \label{eqn:xray}
\frac{\mathrm{d}L_{X}/{\dot{M}_\star}}{\mathrm{d}E} = \frac{E^{-1}}{\int_{\nuX}^{2\,\mathrm{keV}} \mathrm{d}E \, E^{-1}} \times 
\begin{cases}
    L^\textsc{ii}_{X,<2\mathrm{keV}}/{\dot{M}_\star} & \mathrm{PopII}\\
    L^\textsc{iii}_{X,<2\mathrm{keV}}/{\dot{M}_\star} & \mathrm{PopIII}
\end{cases}
\EE 
where $L^\textsc{ii}_{X,<2\mathrm{keV}}/{\dot{M}_\star}$ and $L^\textsc{iii}_{X,<2\mathrm{keV}}/{\dot{M}_\star}$ are the total soft-band X-ray luminosity per unit SFR in atomic cooling and molecular cooling galaxies respectively. In the following we refer to these as $L^\textsc{ii}_X/{\dot{M}_\star}$ and $L^\textsc{iii}_X/{\dot{M}_\star}$ and assume they have the same value $=\LX$, motivated by simulations of high mass X-ray binaries in metal-poor environments in the early universe \citep{Fragos2013} -- though it is possible to vary them independently. \nuX is the cutoff in X-ray energies which can escape galaxies (e.g. X-rays with energy $<\nuX$ are absorbed by the ISM of the host galaxies and thus do not interact with the IGM \citep[e.g.,][]{Das2017}. X-rays at higher energies ($E > 2$\,keV) have a mean free path greater than the Hubble length and thus also do not efficiently heat the IGM.

The astrophysical model described above has a total of 14 free parameters. A summary of these parameters and which ones we vary in our analyses is given in Table~\ref{tab:fiducial_params}.

\subsection{Modelling the 21-cm signal}
\label{sec:method_21cm}

Hydrogen atoms have heavily disallowed hyperfine transitions, key amongst them is the 21-cm line between the singlet and triplet states of the $^1$S orbital. The rate of these `spin-flip' transitions is determined by the relative population of atoms in either states:
\BE
\frac{n_1}{n_0} = \frac{g_1}{g_0}e^{-\frac{T_*}{T_S}}
\EE
where $g_i$ are the spin degeneracy factors of each state (with $g_1/g_0=3$ for 21-cm transitions), $T_*=E_{10}/k_B=68$\,mK corresponding to the transition energy $E_{10}$, and $T_S \gg T_*$ is the \textit{spin temperature}. In the early universe, $T_S$ is coupled to the kinetic temperature of the IGM through collisional excitations~\citep{Loeb:2003ya}, but by $z\sim30$ collisions become rarer and interactions with CMB photons dominate, driving a thermal equilibrium between $T_S$ and $T_\textsc{cmb}$, resulting in no net absorption or emission of 21-cm photons relative to the CMB black-body. However, once the first stars form ($z\sim20-30$), this thermal equilibrium is broken due to the injection of \lya photons \citep[the Wouthuysen–Field effect, e.g.,][]{Hirata2006} and the spin temperature couples to the lower temperature of the IGM gas, resulting in a net absorption of CMB photons at 21-cm wavelengths. Later, as the IGM is heated by X-rays from the first galaxies ($z\sim10-15$), the spin temperature exceeds the CMB temperature resulting in a net emission of 21-cm photons. Once hydrogen is reionized ($z\sim6-10$) the density of neutral hydrogen drops dramatically and the cosmological 21-cm signal is negligible \citep[see e.g.,][for reviews]{Furlanetto2006,Morales2010,Pritchard2012}.

The goal of Cosmic Dawn 21-cm experiments is to measure the 21-cm brightness temperature, \Tb, the offset of the 21-cm signal relative to the temperature of the cosmic microwave background, $T_\textsc{cmb}$ \citep[e.g.,][]{Furlanetto2006}:
\BEA \label{eqn:Tb}
\Tb &=& \frac{T_\textsc{s} - T_\textsc{cmb}}{1+z} (1 - e^{-\tau_{21}})\\ \nonumber
&\approx& 27 \xHI (1 + \delta) \frac{H(z)}{\mathrm{d}v_t/\mathrm{d}r + H(z)} \left(1 - \frac{T_\textsc{cmb}}{T_\textsc{s}}\right) \nonumber \\
&&\times \left(\frac{1+z}{10}\frac{0.15}{\Omega_m h^2}\right)^{1/2} \left(\frac{\Omega_b h^2}{0.023}\right) \mathrm{mK}.
\EEA 
where the second line is the $\tau_{21} \ll 1$ Taylor expansion. Here \xHI is the neutral hydrogen fraction, $T_\mathrm{S}$ is the spin temperature of the gas, $\delta = \rho/\overline{\rho} - 1$ is the overdensity of the gas, and $\mathrm{d}v_t/\mathrm{d}r$ is the gradient of the line-of-sight velocity component. All quantities are computed at $z = \nu_0/\nu - 1$, where $\nu_0 = 1420$\,MHz is the 21-cm rest-frame frequency. This Taylor expansion is useful for physical intuition, though we note that {\tt 21cmFAST} computes the full $\exp{(-\tau_{21})}$ term.

To create the neutral fraction, spin temperature, density and velocity fields required to find \Tb we use the public code {\tt 21cmFAST} \citep{Mesinger2011,Park2019,Murray2020} using the astrophysical model described above in Section~\ref{sec:method_astro} and we refer the reader to these works and references therein. In Section~\ref{sec:method_21cmsense} we describe the parameters and setup for our {\tt 21cmFAST} simulations.

\subsection{Fisher-matrix formalism}
\label{sec:method_fisher}

Given a posterior distribution $P(\mathbf{\theta} | \mathrm{data})$ for model parameters $\mathbf{\theta}$, the Fisher-matrix components are given by:
\BE \label{eqn:fish1}
\mathcal{F}_{ij} = - \left\langle \frac{\partial^2 \ln{P}}{\partial \theta_i \partial \theta_j} \right\rangle
\EE
The Cram\'{e}r-Rao theorem states that any unbiased estimator for the parameters will produce a covariance matrix that is no more accurate than $\mathcal{F}^{-1}$: thus the Fisher matrix can be used to estimate the minimum uncertainties of parameters given observations \citep[e.g.,][]{Albrecht2009}.

Here we focus on the 21-cm power spectrum but the same analysis can be done with the 21-cm global signal \citep[e.g.,][]{Liu2013,Munoz2020}, or any other observable. Previous works have used Fisher-matrix analyses of simulated 21-cm observations with {\tt 21cmFAST} to forecast constraints, for example, on reionization and cosmological parameters, and small-scale dark-matter structure \citep[e.g.,][]{Pober2014,EwallWice2016,Liu2016,Shimabukuro2017,Munoz2020,Shaw2020,Greig2022}. However, these works all used a restricted parameter space, with most using the 3 parameter astrophysical model of {\tt 21cmFAST v1}. Here, we will use the extended {\tt 21cmFAST v3} parameter space (with 14 total degrees of freedom), which characterizes the UV and X-ray emission from both population II and III star formation, as detailed in Section \ref{sec:method_astro}.

Given a list of parameters $\theta_i$, and assuming the posterior distribution can be described as Gaussian (i.e. $\ln{P} \propto -\chi^2$), we define the Fisher matrix for the 21-cm power spectrum $\Delta^2_{21}(k,z)$ as
\BE \label{eqn:Fij_PS}
    \mathcal{F}_{ij} = \dfrac{\partial^2 \chi^2}{\partial \theta_i\partial \theta_j} = \sum_{i_k,i_z} \dfrac{\partial \Delta^2_{21}(k,z)}{\partial{\theta_i}} \dfrac{\partial \Delta^2_{21}(k,z)}{\partial{\theta_j}} \dfrac{1}{\sigma_{\Delta^2}^2(k,z)}
\EE
where $\sigma_{\Delta^2}^2$ is the measurement error in the power spectrum with $k$ bin $i_k$ and $z$ bin $i_z$, and we have assumed uncorrelated errors between $k$ and $z$ bins \citep[in principle we can compute the full measurement covariance matrix but we follow the assumption of uncorrelated errors to more easily compare to previously published MCMC results, e.g.,][]{Park2019}. Our code {\tt 21cmfish} calculates the derivatives with respect to a set of input parameters to produce the Fisher matrix (Equation~\ref{eqn:Fij_PS}).

The inverse of the Fisher matrix is the covariance matrix $\mathcal{C}=\mathcal{F}^{-1}$, and with this definition it is easy to see that the forecasted uncertainty in the $i$-th parameter is simply $\sigma(\theta_i) = \sqrt{\mathcal{C}_{ii}}$.

\subsection{Simulated 21-cm observations and uncertainties}
\label{sec:method_21cmsense}

\begin{table*}
\caption{21cmFAST parameters and their fiducial values for our two runs. For a detailed description see Sec.~\ref{sec:method_astro}. }             
\label{tab:fiducial_params}      
\centering       
\begin{tabular}{ lllrcrc }
\hline\hline
\multicolumn{3}{c}{} & \multicolumn{2}{l}{\citet{Park2019}} & \multicolumn{2}{l}{EOS21 \citep{Munoz2021b}}\\ \cline{4-5} \cline{6-7}
Parameter & Description & Population & Fiducial & Vary? & Fiducial & Vary?\\
\hline\hline
\astarII & 
Stellar -- halo mass power law slope & 
II &
$0.50$ & \checkmark &
$0.50 $ & \checkmark \\
$\log_{10}\fstarII$ & Stellar -- halo mass normalization (at $M_h=10^{10} M_\odot$) & 
II &
$-1.30$ & \checkmark &
$-1.25$ & \checkmark  \\
\tstar & SFR timescale as a fraction of the Hubble time&
both &
$0.50$ & \checkmark &
$0.50$ &  \\
$\log_{10}\Mturn$ & Halo mass turnover for atomic cooling halos [$\mathrm{M}_\odot$]& 
II &
$8.7$ & \checkmark &
 -- & \\
\astarIII &  Stellar -- halo mass power law slope &
III &
 -- & &
$0.0$ & \checkmark \\
\fstarIII & Stellar -- halo mass normalization (at $M_h=10^{7} M_\odot$) &
III &
-- & &
$-2.50$ & \checkmark  \\
\hline
\aesc & Ionizing escape fraction -- halo mass power law slope & 
both &
$-0.50 $ & \checkmark &
$-0.30 $ & \checkmark \\
$\log_{10}\fescII$ & Ionizing escape fraction normalization (at $M_h=10^{10} M_\odot$) & 
II &
$-1.00$ & \checkmark &
$-1.35$ & \checkmark \\
$\log_{10}\fescIII$ & Ionizing escape fraction normalization (at $M_h=10^{7} M_\odot$) &  
III &
-- &  &
$-1.35$ &  \checkmark   \\
\hline
$\log_{10}L^\textsc{ii}_X/{\dot{M}_\star}$ & X-ray luminosity per SFR [erg s$^{-1}$ M$_\odot^{-1}$ yr] &
II &
$40.5$ & \checkmark  &
$40.5$ & \checkmark  \\
$\log_{10}L^\textsc{iii}_X/{\dot{M}_\star}$ & X-ray luminosity per SFR [erg s$^{-1}$ M$_\odot^{-1}$ yr] & 
III &
-- &   &
$40.5$ & $=\log_{10}L_X^\textsc{ii}$  \\
\nuX & minimum X-ray energy which escapes galaxies [eV] &
both &
$500$ & \checkmark &
$500$ & \checkmark \\
\hline
$A_{v_\mathrm{cb}}$ &  Amplitude of DM-baryon relative velocity feedback & 
III &
-- & &
$1.00$ &  \\
$A_{LW}$ & Amplitude of Lyman-Werner feedback & 
III &
-- & &
$2.00$ & \checkmark \\
\hline
\end{tabular}
\end{table*}

As described in Section~\ref{sec:method_fisher} above, we can use the Fisher matrix to derive the lowest possible estimate of the parameter uncertainties given observations. In this work we generate parameter constraints from two sets of mock observations, each serving a specific purpose:
\begin{enumerate}
    \item We validate our Fisher-matrix approach by comparing to the MCMC posterior of \citet{Park2019}. We use their fiducial parameters, which only includes population II star formation in atomic-cooling galaxies. Simulation boxes have a comoving volume of (250\,Mpc)$^3$ on a $128^3$ grid from $z=5.9-28$ following \citet{Park2019} to match the HERA observing bandwidth.
    \item An exploration of the updated `Evolution of Structure` (EOS21) simulation presented by \citet{Munoz2021b}\footnote{Simulation data available at: \url{https://www.dropbox.com/sh/dqh9r6wb0s68jfo/AACc9ZCqsN0SQ_JJN7GRVuqDa?dl=0}}. The EOS21 parameters include star formation in minihalos, and the impact of Lyman-Werner feedback and relative velocities between baryons and dark matter. The PopII parameters are based on those inferred by \citet{Qin2021b}, who used the \citet{PlanckCollaboration2016} electron scattering optical depth, rest-frame UV luminosity functions \citep[LFs,][]{Bouwens2015b}, dark pixel fraction \citep{McGreer2015}, and \lya\ forest optical depth \citep{Bosman2018} in their likelihood.
    On the other hand, the PopIII parameters are
    difficult to constrain with current observations, and so were chosen fairly arbitrarily.  We note PopIII star formation in their fiducial model only dominates the total SFRD at $z\gtrsim 15$.
    The simulation boxes we use have a comoving volume of (400\,Mpc)$^3$ on a $200^3$ grid from $z=5-30$. 
\end{enumerate}

The fiducial parameters for these runs are given in Table~\ref{tab:fiducial_params}. Our motivation for choosing these sets of fiducial parameters is that they produce UV LFs, a reionization history and CMB scattering optical depth that are consistent with current observations at $z\sim6-10$, and so are thus likely to represent a reasonable parameter space for high redshift galaxy properties. We use $\geq (250$\,Mpc)$^3$ to ensure convergence of the power spectrum in the range $\sim0.1-1$\,Mpc$^{-1}$ which we will consider below, following \citet{Kaur2020}. Our resolution of $\approx 2$\,Mpc per voxel should also provide a converged power spectrum: \citet{Greig2022} demonstrated that this resolution, with a similar volume to our simulations, provides converged wavelet scattering transformation coefficients (a non-Gaussian probe, complementary to the power spectrum) for the 21-cm signal.

\begin{figure*}
\includegraphics[width=0.49\textwidth]{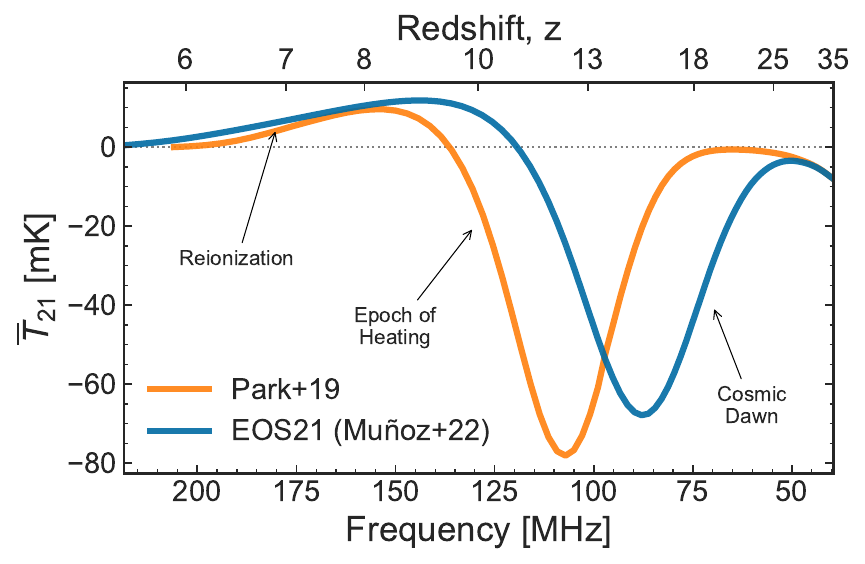}
\includegraphics[width=0.49\textwidth]{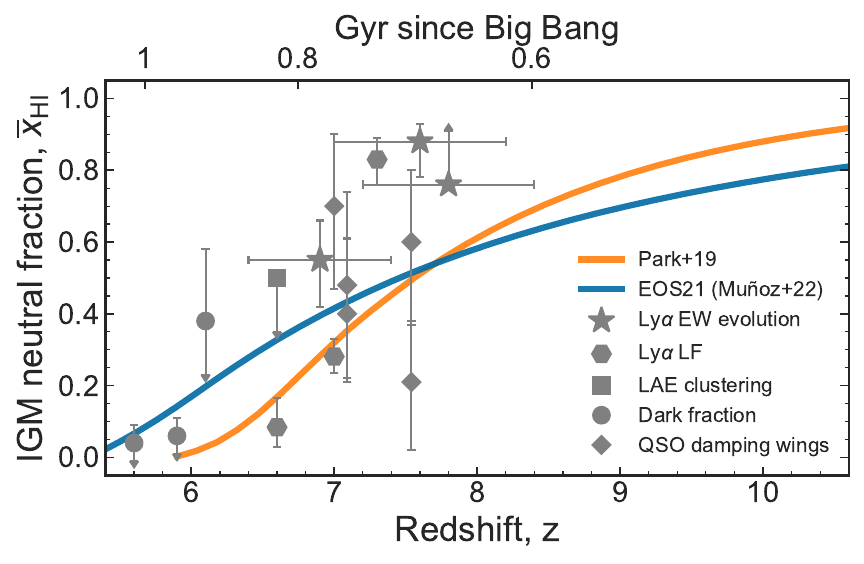}
\includegraphics[width=\textwidth]{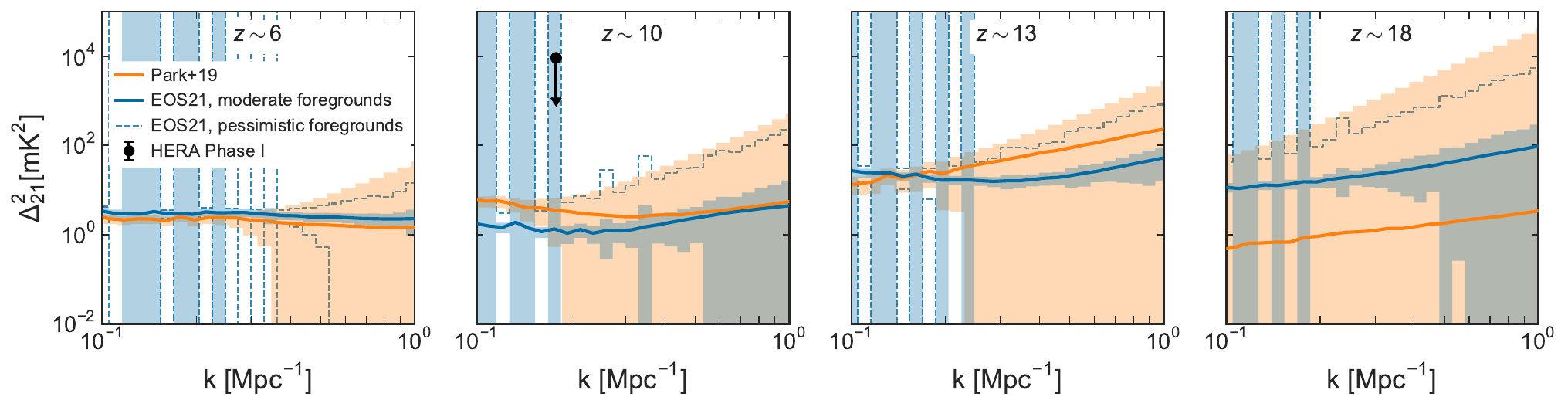}
\caption{Comparison of our two fiducial models, with \citet{Park2019} in orange, EOS21 in blue. \textbf{Top left:} Global 21-cm signal. \textbf{Top right:} Reionization history. We also plot reionization history constraints derived from observations of the \lya equivalent width distribution \citep[star,][]{Mason2018,Mason2019,Whitler2020, Hoag2019a}; the \lya luminosity function \citep[hexagon,][]{Morales2021} the clustering of \lya emitting galaxies \citep[square,][]{Ouchi2010,Sobacchi2015}; \lya and Ly$\beta$ forest dark pixel fraction \citep[circle;][]{McGreer2015}; and QSO damping wings \citep[diamond,][]{Greig2017,Davies2018,Greig2019,Wang2020}. \textbf{Lower:} Example power spectra of our two fiducial models at four redshifts. We show our fiducial `moderate' HERA noise estimates for each case as the filled regions, and the `pessimistic' noise as dashed blue line, including Poisson noise and a 20\% modelling error added in quadrature. We show the errors only in the $k$-space window where we calculate the likelihood ($0.1-1$\Mpcinv). Note that due to a spline interpolation of the power spectra errors by \citet{Park2019}, the low $k$ noise estimates are smooth, unlike those in the EOS21 model. For comparison, we also plot the 95\% confidence upper limit at $z=10.4$ from HERA Phase I observations \citep{HERA2021a}.
\label{fig:fiducials}
}
\end{figure*}

21-cm power spectra are generated by dividing lightcones into `chunks' following \citet{Greig2018}. For our comparisons with \citet{Park2019} we follow their approach and divide the lightcones into 12 chunks of equal comoving volume. For our EOS21 run we divide the lightcone into chunks corresponding to 8\,MHz bandwidths, which more closely resembles what would be done with data. The 21-cm global signal, reionization history and power spectra at a few redshifts for the two models are plotted in Figure~\ref{fig:fiducials}.

To calculate the Fisher matrices we vary each parameter $\pm3\%$ of its fiducial value and calculate two-sided derivatives\footnote{The variation is in either linear or logarithmic quantities corresponding to the parameters in Table~\ref{tab:fiducial_params}, i.e. we vary $\log_{10}\fstarII \pm3\%$ of its fiducial value.}, finding the variation step-size such that the derivatives were converged. We tested convergence of the derivatives and found $\pm3\%$ worked in all cases with the exception of $\log_{10}\LX$ which we varied by $\pm0.1\%$ and $\log_{10}\Mturn$ which we vary by $\pm1\%$, as these parameters have narrower likelihoods in the 21-cm signal. We verified that one-sided derivatives produced almost identical values.

Thus, to create a derivative $\partial \Delta^2_{21}(k,z)/\partial{\theta_i}$ for each parameter we use 3 simulated power spectra: the fiducial, one setting the parameter to $\theta_i(1+X)$ and one with $\theta_i(1-X)$, where $X$ is the variation (usually 3\% as described). As we only need to create one fiducial simulation, for $N$ parameters we create a total of $2N + 1$ simulations to calculate the Fisher matrix (Equation~\ref{eqn:Fij_PS}).

To make forecasts for upcoming observations with HERA we use the python package {\tt 21cmSense}\footnote{\url{https://github.com/jpober/21cmSense}} \citep[][]{Pober2014}. We run three different setups, which we now describe. For the comparison with \citet{Park2019} we use the same noise as their study, which is essentially identical to the `pessimistic' case described below. The noise was calculated assuming 1000 hours of observation using 331 antennae.

For our EOS21 forecasts (with PopIII stars in molecular cooling galaxies) we find the noise in two cases, with moderate and pessimistic foregrounds, respectively. The dominant sources of noise for 21-cm observations are the instrumental noise of the antennae and large foreground contamination, from both our Galaxy and the atmosphere. The combination of these noise sources can be parameterised by a ``system temperature'', $T_{\rm sys}(\nu)$ which is the total observed 21-cm brightness temperature \citep[e.g.][]{Morales2010}. The foregrounds substantially contaminate the signal in the Fourier plane, particularly in the so-called ``wedge'' \citep[e.g.,][]{Liu2009,Pober2013,Liu2020}, within wavenumbers $k_\parallel \leq a + b k_\bot$. Here $k_\parallel$ and $k_\bot$ are the wavenumbers in the line-of-sight and perpendicular direction, and $a$ (called the super-horizon buffer in {\tt 21cmSense}) and $b$ are constants that parameterise the severity of the foregrounds and thus which regions of $k$-space will be discarded in the analysis.

In the moderate foregrounds case we use the `moderate' foregrounds model in {\tt 21cmsense}, with a wedge super-horizon buffer $a=0.1\,h \Mpcinv$ and a HERA system temperature \citep{deBoer2017} of
\begin{equation} \label{eq:TsysHERA}
    T_{\rm sys} (\nu) = 100\,\mathrm K + 120 \,\mathrm K \left(\nu/150\rm MHz \right)^{-2.55}.
\end{equation}
In the pessimistic case we increase the wedge horizon buffer to $a=0.15\,h \Mpcinv$, and use the default system temperature in {\tt 21cmSense}, which is $\sim 3\times$ larger than that of Equation~\eqref{eq:TsysHERA} \citep{Pober2014}. In both cases, the antennae temperature is the same but this larger system temperature is motivated by the sky temperature measured by LOFAR \citep{vanHaarlem2013}. In the pessimistic case the signal-to-noise ratio of the power spectrum is lower by nearly an order of magnitude, and consequently increases the error bars on each parameter by approximately a factor of three, as we will see (Figures~\ref{fig:corner_EOS21_noise} and \ref{fig:EOS21_SNR}). This pessimistic case is comparable to that used in the forecasts by \citet{Park2019}. In both moderate and pessimistic cases for EOS21 we assume 1 year (1080 hours) of data using 331 antennae across the 50-250 MHz range (corresponding to the redshifts $z\sim5-28$), divided in equal-width bands of 8\,MHz.

In all cases, we follow \citet{Park2019} and add Poisson noise (from the finite-size simulations) and a 20\% modelling error to the power spectra in quadrature in addition to the Poisson noise for the mock observations. We calculate the likelihood only in the $k$-space window $0.1-1$\Mpcinv, corresponding roughly to limits on the foreground noise and the shot noise, respectively. For the \citet{Park2019} comparison we calculate the likelihood over the redshift range $z=5.9-28.0$ to match their analysis, and for EOS21 we calculate the likelihood over $z=5.0-28.0$

\section{Results} 
\label{sec:results}

Section~\ref{sec:res_Park19} describes the validation of our Fisher-matrix approach by comparison to an MCMC by \citet{Park2019}. Section~\ref{sec:res_full} presents forecasted constraints on the  population II and population III galaxy formation parameter space expected with HERA 1 year observations, assuming the EOS21 model from \citet{Munoz2021b}. Section~\ref{sec:res_full_PCA} presents the principal components of the Fisher matrix, providing insight into the best constrained linear combinations of parameters. In Section~\ref{sec:res_inference} we evaluate whether the Fisher matrix can be used for inference.

\subsection{Comparison with \citet{Park2019} population II-only MCMC}
\label{sec:res_Park19}

\begin{figure*}
\includegraphics[width=\textwidth]{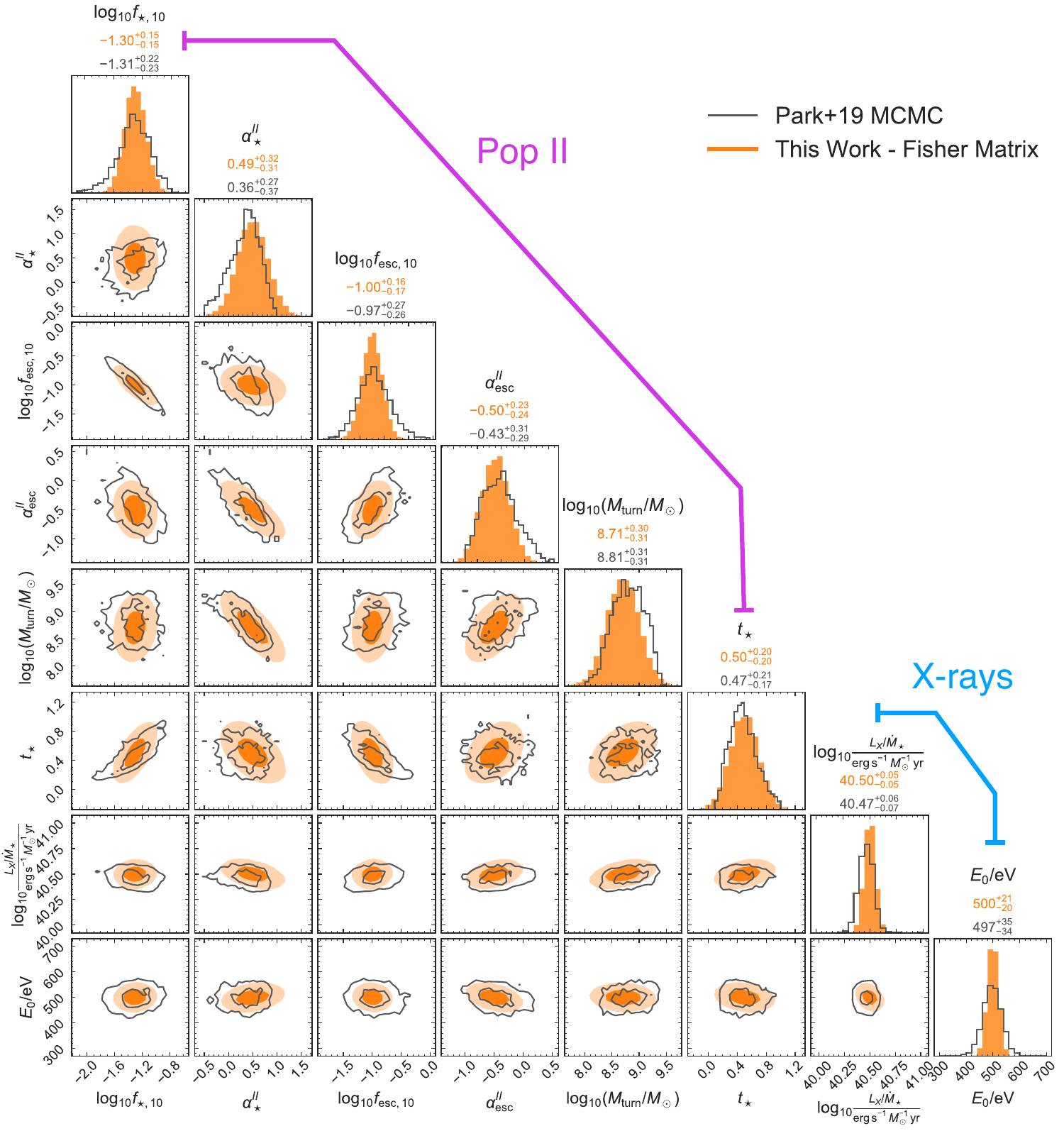}
\caption{Validation of our Fisher-matrix technique by comparison with the MCMC runs of \citet{Park2019}. We plot parameter constraints from \citet{Park2019} (grey line) and our results (orange filled). The top right plots show 1D marginalized posteriors and the lower triangle plots show the 2D marginalized posteriors. The contours show $1,2\sigma$ confidence intervals for the posteriors. Our Fisher-matrix approach recovers the same degeneracies as the MCMC and comparable uncertainties, while requiring $\sim 10^4$ fewer simulations.
\label{fig:corner_Park19}
}
\end{figure*}

To verify our Fisher-matrix approach we compare with the MCMC analysis by \citet{Park2019}. \citet{Park2019} carried out an inference on mock HERA observations to infer properties of population II galaxies. The parameters varied in this analysis are listed in Table~\ref{tab:fiducial_params}.

To compare with \citet{Park2019} we construct simulations using the same dimensions and redshift range, as described in Section~\ref{sec:method_21cm}, and with the same fiducial parameters, though with different initial conditions. We then divide lightcones into 12 chunks of equal comoving volume to create power spectra, replicating their approach, and calculate the derivatives of the power spectrum with respect to the parameters of interest as described in Section~\ref{sec:method_fisher}. To create the Fisher matrix (Equation~\ref{eqn:Fij_PS}) we use the same HERA noise estimate $\sigma[\Delta^2(k,z)]$ as \citet{Park2019}, which was obtained using {\tt 21cmsense}. To obtain the total uncertainty on the power spectrum, we add in quadrature the HERA thermal noise plus the cosmic variance plus a 20\% `modelling uncertainty' on the power spectrum, again following \citet{Park2019}.

The resulting $1,2\sigma$ error ellipses obtained using our Fisher matrix are shown in Figure~\ref{fig:corner_Park19}, along with the 21cm-only\footnote{\citet{Park2019} also shows posteriors obtained by combining existing UV LF observations with the mock 21cm power spectra in the likelihood.} MCMC posteriors by \citet{Park2019}. We note that we plot the MCMC chains produced by \citet{Park2019}, but do not exactly reproduce their published posteriors and 68\% credible intervals due to some small differences in burn-in steps and removal of local minima in the posterior between that work and ours. The chains have 229 steps and 35 walkers. We remove the first 50\% of steps as burn-in and also remove $\sim24\%$ of the chains which we identify as falling into local minima. The median maximum likelihood for the MCMC was $\mathcal{L}_\mathrm{max} \sim 10^{-4}$ and we remove chains in local minima -- where the likelihood in the final step is $< 10^{-10}$.

Our Fisher-matrix approach clearly recovers the inferred degeneracies between parameters in all cases, and produces comparable uncertainties to those estimated using the MCMC. The MCMC contours and the Fisher ellipses are oriented similarly and have similar ellipticities demonstrating that the correlations between parameters are well-captured by the Fisher matrix analysis. The marginalised 68\% credible intervals on each parameter obtained with the Fisher matrix are within $40\%$ of those obtained with the MCMC (with the exception of \nuX where the Fisher error is $\sim50\%$ of the MCMC value). These results do not change significantly if we use the median inferred parameters obtain by \citet{Park2019}, implying any difference in uncertainties are dominated by the approximation of the posterior as Gaussian, rather than the position of the posterior peak. As noted in Section~\ref{sec:method_fisher}, the Fisher matrix will give the minimum uncertainties, so we do expect our technique to slightly underestimate the errors compared to a full mapping of the posterior. Nevertheless, this remarkable agreement validates our Fisher matrix approach as an efficient tool to explore 21-cm observational constraints on cosmology and astrophysics, providing the parameter degeneracies remain approximately Gaussian, dramatically reducing the need for expensive MCMC runs. For comparison, the MCMC run by \citet{Park2019} required 70,000 individual simulations, whereas our Fisher-matrix approach required only 17.

The parameter where the Fisher-matrix constraints are most different from the MCMC is \astarII. This is because \citet{Park2019} set a uniform prior on $\astarII = [-0.5,1]$, whereas in the Fisher-matrix approach we do not set any bounding conditions and make the assumption of a Gaussian likelihood, so the confidence ellipse can extend to $\astarII > 1$, making it broader than the MCMC posterior.

Due to using the `pessimistic' foreground model assumed (see Section~\ref{sec:method_21cmsense}), these forecasts find most parameters will be measured at the $20-30\%$ error level, with the exception of X-ray parameters and $\Mturn$ which can be measured to $<10\%$ precision, whereas $\astarII$ and $\aesc$ are more difficult to measure precisely. As discussed by \citet{Park2019}, the inclusion of independent observations that can constrain galaxy formation properties, for example the UV luminosity function, can break degeneracies and reduce parameter uncertainties to $\lesssim10\%$ precision.

\subsection{Extended parameter space forecast}
\label{sec:res_full}

\begin{figure*}
\includegraphics[width=\textwidth]{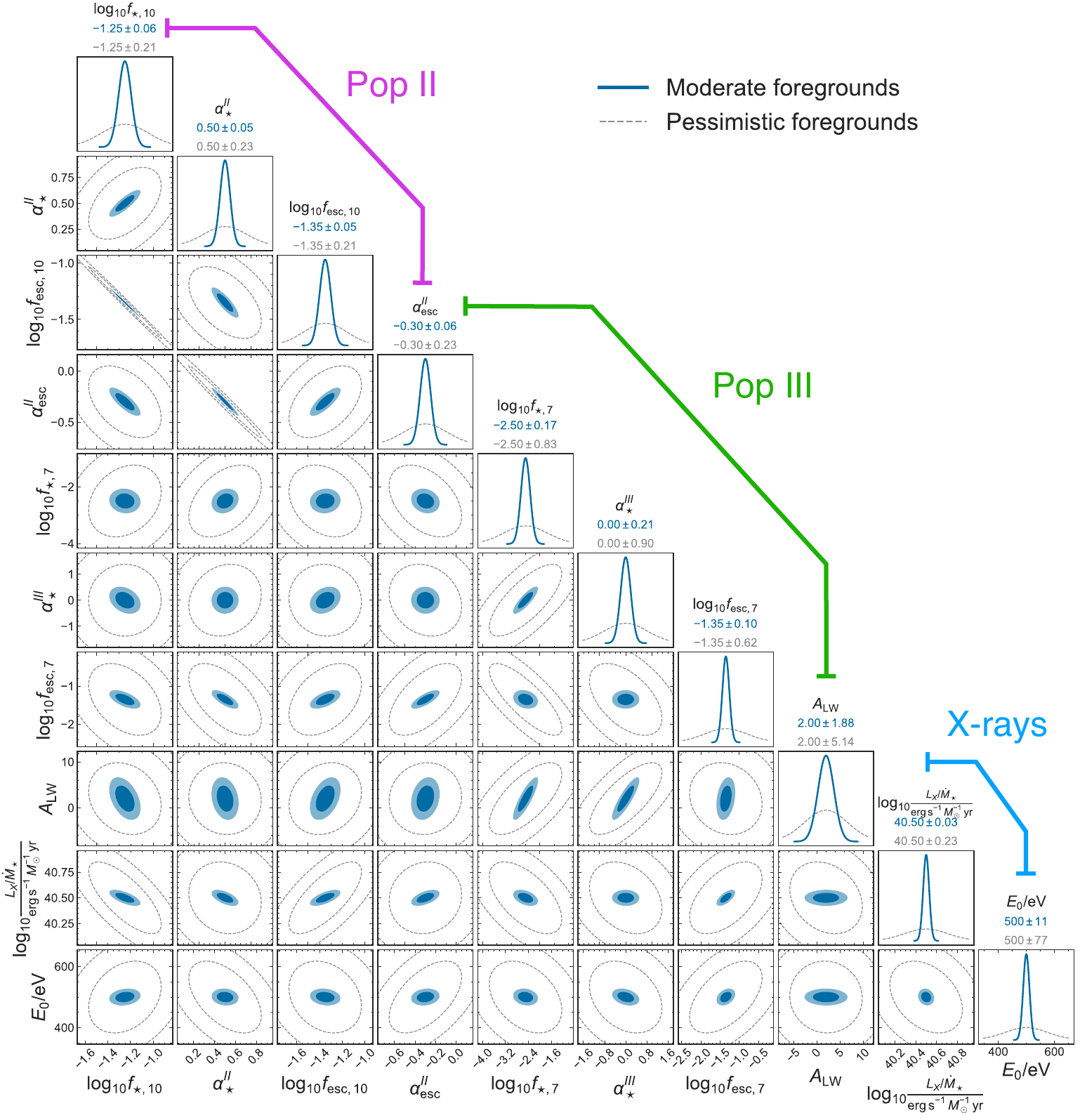}
\caption{Parameter constraints for the fiducial EOS21 run, using the parameters defined by \citet{Munoz2021b} for moderate (blue solid) and pessimistic (grey dashed) foregrounds in HERA 1-year observations, as described in Section~\ref{sec:method_21cmsense}. The contours show 1 and 2$\sigma$ confidence intervals and the quoted confidence interval for the 1D constraint is 1$\sigma$.
\label{fig:corner_EOS21_noise}
}
\end{figure*}

\begin{figure}
\includegraphics[width=\columnwidth]{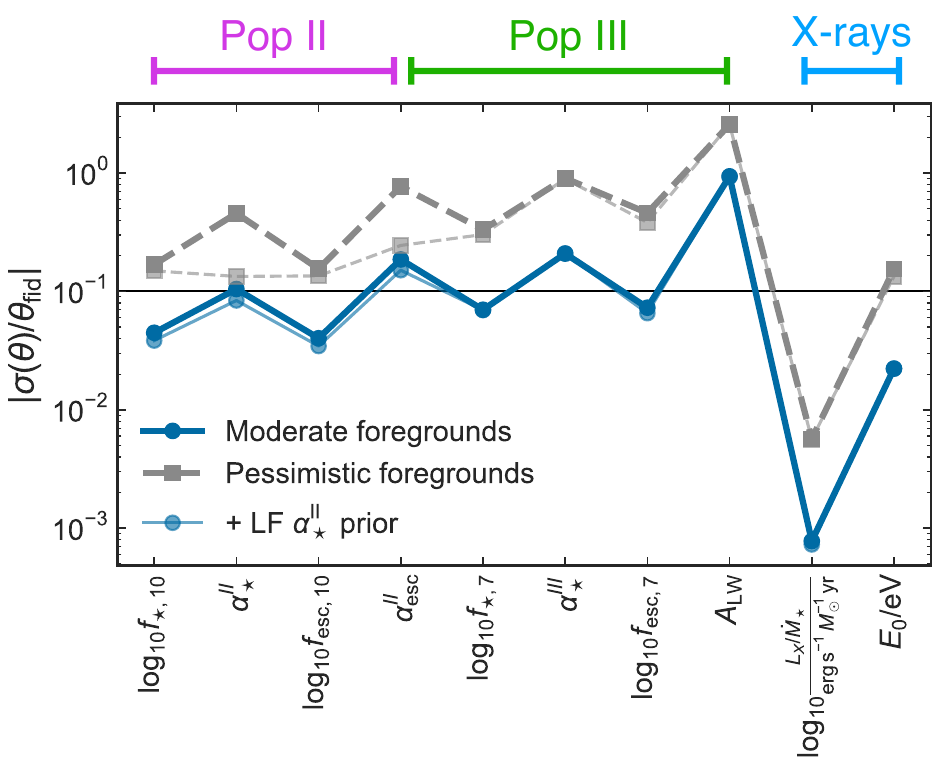}
\caption{Fractional $1\sigma$ error in each parameter as estimated from the Fisher matrix, for the moderate (blue solid line) and pessimistic (grey dashed line) foreground noise models (see Section~\ref{sec:method_21cmsense}). In both cases, the thick lines show the HERA-only constraints and the thin solid lines in the same color and style show the fractional error estimate after including a prior on \astarII from the UV LF using Hubble data. The \astarII prior reduces the uncertainty on population II parameters by approximately a factor of three. We mark the 10\% fractional uncertainty with a solid horizontal line for ease of comparison.
\label{fig:frac_error}
}
\end{figure}

\begin{figure}
\includegraphics[width=\columnwidth]{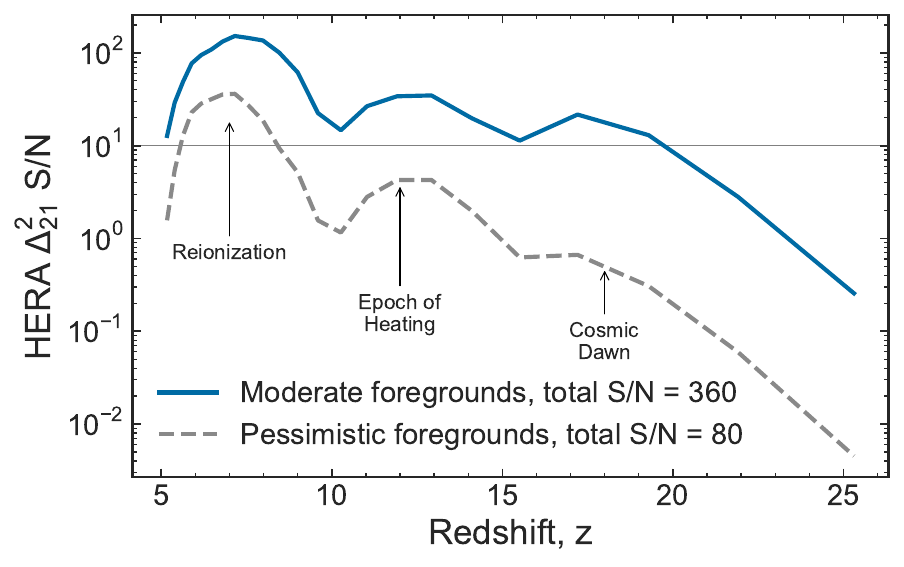}
\caption{Signal to noise ratio of our fiducial EOS21 21-cm power spectrum as a function of redshift, using the moderate (blue solid) and pessimistic (grey dashed) expected noise from 1 year of HERA data, as described in Section~\ref{sec:method_21cmsense}. Due to the frequency/redshift dependence of the noise, the pessimistic noise level would only enable high S/N measurements during the Epoch of Reionization, thus constraints on population III galaxies would be very limited, as seen in Figure~\ref{fig:corner_EOS21_noise}.
\label{fig:EOS21_SNR}
}
\end{figure}

\begin{figure}
\includegraphics[width=\columnwidth]{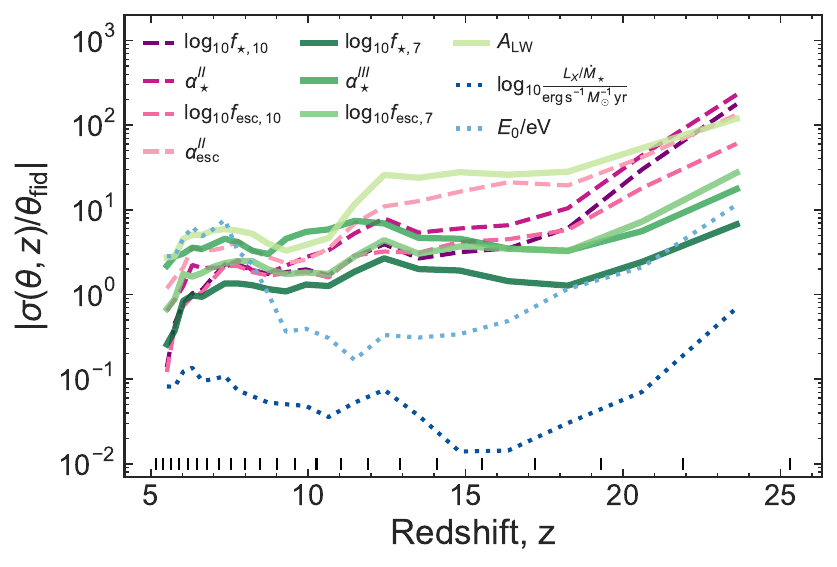}
\caption{Fractional $1\sigma$ error in each parameter as estimated from the Fisher matrix in rolling redshift bins corresponding to two adjacent HERA frequency channels, demonstrating the redshift evolution of measurable constraints on each parameter. All lines show the constraints using the moderate foreground noise model (see Section~\ref{sec:method_21cmsense}). Pink dashed lines show population II parameters, green solid lines show population III parameters, and blue dotted lines show X-ray parameters. Vertical black ticks show the redshift bin edges for our mock HERA observations.
\label{fig:frac_error_z}
}
\end{figure}

Given that our Fisher-matrix approach recovers the same degeneracies and very similar confidence intervals as an MCMC, we now make a forecast in an unexplored regime: 10 free astrophysical parameters governing both population II and population III galaxy formation, and the impact of streaming velocities and Lyman-Werner feedback on the formation of minihalos, over the full redshift range of the Cosmic Dawn 21-cm signal ($z \sim 5-30$). Due to computational expense and current instrumental sensitivity, previous works have performed MCMC or Nested sampling in a more limited parameter space, varying only PopII or PopIII parameters at a time \citep[e.g.,][explored 7 and 9 parameters respectively, roughly requiring $\sim10^5$ individual simulations and $\sim$100k CPUh]{Qin2021a,HERA2021b}. Our {\tt 21cmfish} calculation, by contrast, required only 21 individual simulations (as we fixed $L_X^\textsc{ii}/\dot{M}_\star = L_X^\textsc{iii}/\dot{M}_\star$). Although the Fisher matrix assumes a multi-variate Gaussian posterior and the setting of physical priors is less straightforward, these are reasonable approximations when the posterior is uni-modal and narrowly peaked around the maximum likelihood value (e.g. \citealt{Trotta2008}). Indeed, we expect future HERA and SKA observations to be highly constraining, making Fisher forecasts useful approximations of the true posterior for futuristic data sets.

Figure~\ref{fig:corner_EOS21_noise} shows the forecasted parameter constraints from our Fisher-matrix approach for 10 EOS21 fiducial free parameters \citep{Munoz2021b}. The contours show $1,2\sigma$ confidence intervals for the parameters. For this forecast, we assume one year (i.e., 1080 hours) of HERA data, and show the forecasts for moderate and pessimistic foreground noise levels, as described in Section~\ref{sec:method_21cmsense}. 

The fractional error estimated for each parameter is plotted in Figure~\ref{fig:frac_error}. With moderate foreground noise it would be possible to infer nearly all parameters to $\simlt10-30\%$ precision, with the exception of the amplitude of Lyman-Werner feedback, $A_\mathrm{LW}$, which is much more poorly constrained -- due to its smaller impact on the amplitude of the 21-cm signal \citep{Munoz2021b}, and X-ray parameters which can be constrained to $<10\%$. This demonstrates 21-cm observations will be a powerful tool to learn about the formation of the first galaxies. As expected, parameter constraints would be substantially worse in the case of `pessimistic' foreground noise, particularly on population III parameters. This is because the foreground noise is a strong function of frequency and thus redshift, with the lowest noise in the $z<10$ -- Epoch of Reionization -- window, which is dominated by population II-hosting galaxies. This can be seen clearly in the signal-to-noise ratio of our fiducial 21-cm power spectrum as a function of redshift for the two noise models, which we plot in Figure~\ref{fig:EOS21_SNR}.

We find that most parameters for \textit{both} population II and III galaxies could be measured at $<10\%$ precision under the assumption of moderate instrumental noise. This is due to the relative redshift dependence of the parameters and the wide frequency/redshift range of the HERA observations. In Figure~\ref{fig:frac_error_z} we plot the redshift dependence of the fractional error on each parameter, calculated in rolling redshift bins corresponding to two adjacent HERA frequency 8 MHz bandwiths. From this figure we see the lowest errors on population II parameters are obtained during the later stages of the Epoch of Reionization, $z\sim5-7$ (see Figures~\ref{fig:frac_error_z} and \ref{fig:PC_sigma_z}), when they strongly affect the 21-cm signal. At these redshifts the observational noise is lowest (Figure~\ref{fig:EOS21_SNR}) and the redshift resolution due to our sampling of the lightcone in fixed frequency bins is highest, which will naturally increase the signal-to-noise for population II parameters. But by $z\simgt10$ the errors on population II parameters become comparable to the errors on population III parameters, though the measurements are obtained in sparser redshift bins, and by $z\simgt15$ the errors on population III parameters are lower than the population II parameters -- as lower mass galaxies start to dominate the star formation rate density and thus the 21-cm signal \citep[e.g., see Figure 6 by][]{Munoz2021b}. Due to the wide frequency range of HERA, the redshift dependence of the uncertainties of the parameters is captured, thus the population III parameters can be well-constrained. We can also see from Figure~\ref{fig:frac_error_z} that, as expected, the lowest errors on X-ray parameters are obtained during the Epoch of Heating ($z\sim10-18$ for the EOS21 model).

The Fisher forecast demonstrates there are strong degeneracies between the normalization and low-mass slope of the stellar-halo mass relation for both population II and III galaxies. There is also a very strong degeneracy between \fstarII and \fescII as noted by \citet{Park2019}. \LX is most strongly degenerate with population II parameters, but not as strongly with population III parameters. This stronger degeneracy is likely because population II galaxies are expected to dominate the star formation rate density during the Epoch of Heating \citep[$z\simlt15$, see e.g.,][]{Munoz2021b} and so the relative abundance of these galaxies is thus degenerate with their X-ray emission, whereas population III galaxies dominate the SFR density at higher redshifts, where X-ray heating has a smaller effect on the 21-cm signal.

In terms of an interplay between population II and III parameters, the strongest degeneracies are between \fescIII and population II parameters (\astarII, \fstarII, \aesc, \fescII), attributed to the uncertainty of the relative role of population III-hosting galaxies in reionization. Population III star formation parameters are also degenerate with $A_\mathrm{LW}$, which is expected as it sets the strength of Lyman-Werner feedback which suppresses star formation in molecular cooling halos.

We also produce forecasts for HERA observations adding a prior on $\astarII$. \citet{Park2019} demonstrated that adding other astrophysical observations to the 21-cm signal, in their case the UV luminosity function (LF), provides independent constraints on the model parameters. In their case they found uncertainties in population II star formation: $\fstarII$ and $\astarII$, were reduced by a factor of three when adding UV LF data to their likelihood, as the UV LF provides important information to constrain the stellar-to-halo mass relation (Equation~\ref{eqn:mstell_mhalo}). Here we use the LF-only constraint on $\astarII$ inferred by \citet{Park2019} from Hubble Space Telescope observations, $\sigma(\astarII) = 0.07$, as a prior, by adding $1/\sigma^2(\astarII)$ to the diagonal element corresponding to $\astarII$ in the Fisher matrix \citep[e.g.,][]{Coe2009}. The impact on the total uncertainties in each parameter is shown in Figure~\ref{fig:frac_error}. Using the moderate foreground noise model, the impact of the \astarII prior is negligible, as population II parameters can already be measured to similar precision to the prior, but the prior significantly reduces uncertainties on population II parameters in the case of the pessimistic foreground noise.

Our forecasts demonstrate that observations of the 21-cm power spectrum could be able to determine the efficiency of both population II and population III star formation, if the foreground noise follows the `moderate' prescription, and determine the strength of X-ray production by the first galaxies to sub-percent precision even under `pessimistic' noise. However, an understanding of the observational noise level is crucial. We note that although the `pessimistic' foreground model is comparable to the noise model used by \citet{Park2019} our forecasted uncertainties for the EOS21 model can be higher because the fiducial parameters are different, and the increased number of free parameters adds additional uncertainty due to degeneracies between parameters.

\subsection{Principal Components of the 21-cm Power Spectrum}
\label{sec:res_full_PCA}

\begin{figure}
\includegraphics[width=\columnwidth]{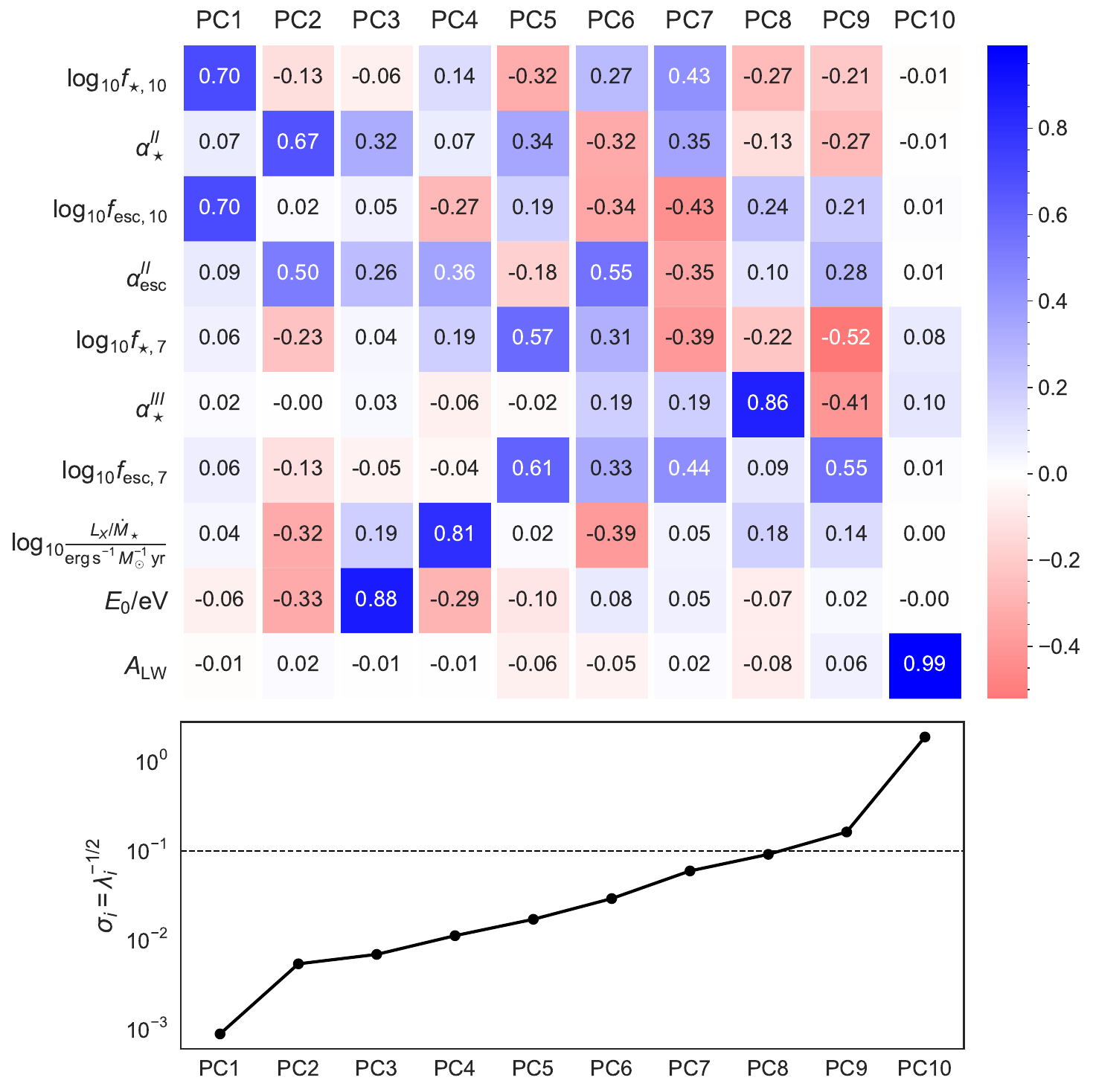}
\caption{(Upper) Eigenvectors associated with each principal component, in order of decreasing importance/increasing uncertainty. For example, PC1 is $\approx \fescII \times \fstarII$. (Lower) Uncertainty for each principal component.
\label{fig:PC_sigma}
}
\end{figure}

Given the parametric model for galaxy formation in {\tt 21cmFAST}, we may ask which combination of parameters can be best constrained by upcoming data. To obtain the most important linear combinations of parameters in the 21-cm signal we carry out a principal component analysis of the Fisher matrix \citep[e.g.,][]{Efstathiou2002,Munshi2006}.

To ensure all parameters are compared at approximately the same order of magnitude we rescale $\nuX$ from eV to keV. Thus, we rescale the Fisher elements corresponding to $\nuX$ by a factor $10^3$.

We can diagonalize our adjusted Fisher matrix $\mathbf{\mathcal{F}^\prime}$:
\BE \label{eqn:PCA_fisher_diag}
\mathbf{\mathcal{F}^\prime} = \mathbf{U} \mathbf{\Lambda} \mathbf{U}^\mathrm{T}
\EE
where $\mathbf{\Lambda}$ is a diagonal matrix with elements $\lambda_i$ and $\mathbf{U}$ is the matrix of principal component eigenvectors, from which we can obtain the orthogonal linear combinations of the parameters $\mathbf{\theta}$:
\BE \label{eqn:PCA_fisher_X}
\mathbf{X} = \mathbf{U}^\mathrm{T} \mathbf{\theta}
\EE
and the variance of the principal component (PC) $X_i$ is $1/\lambda_i$ \citep[e.g.][]{Efstathiou2002}.

Figure~\ref{fig:PC_sigma} shows each PC and its associated uncertainty ($\sigma_i = \lambda_i^{-1/2}$) as a function of the PC number $i$, in order of decreasing accuracy (lowest variance is the first PC). The first 7 out of 10 principal components can be measured to $<10\%$ accuracy, and only one, dominated by $A_\mathrm{LW}$, to $>100\%$.

These principal components demonstrate which combinations of parameters can be best constrained by HERA-like 21-cm power spectra. We obtain the principal components for the entire redshift range of HERA ($z\sim5-28$) and note that the lowest variance components detectable in the full redshift range are mainly set by the frequency range with the lowest noise -- i.e. the PCs that contribute the most at $z\simlt8$ are heavily upweighted due to the much lower noise in that frequency range relative to the Cosmic Dawn window at $z\simgt12$. We explore the PCs in a few redshift windows in Appendix~\ref{app:PCA}.

Several of the PCs have intuitive physical interpretations:

\begin{enumerate}
    \item PC1 is dominated by \fstarII and \fescII. As these two quantities are constrained in log-space, this tells us that the product $\fstarII \times \fescII$ will be well-constrained by HERA. This is unsurprising as this product determines the number of ionizing photons which are available for reionization and these quantities are very degenerate (see Figure~\ref{fig:corner_EOS21_noise}).
    \item PC2 is dominated by a combination of population II and X-ray parameters, as these parameters have a strong impact on the signal during the Epoch of Reionization.
    \item PC3 is dominated by \nuX, which will be easily measured to high precision with minimal degeneracies with other parameters as it determines the hardness of the X-ray spectrum that emerges from early galaxies.  Values of \nuX below $\sim$ 1keV have a strong impact on the inhomogeneity of X-ray heating (and thus the associated 21-cm power spectrum) since the mean free path is a strong function of the photon energy \citep[e.g.,][]{Pacucci2014,Fialkov2016,Das2017}.
    \item PC4 is dominated by \LX, which governs the relative timing of the X-ray heating epoch
    \item Higher principal components have combinations of population II and population III parameters, with the combinations of population III parameters having the highest variance. As discussed in Appendix~\ref{app:PCA} the principal components are dominated by which parameters can be measured in the redshift windows with lowest noise -- thus when looking at the data over the full HERA observing window $z\sim5-28$ population II reionization parameters will be best measured.
    \item PC10 is dominated by $A_\mathrm{LW}$. As discussed in Section~\ref{sec:res_full} this parameter is least constrained by 21-cm power spectra observations as it has a relatively low impact on the signal \citep[see also Figure~17 by][]{Munoz2021b}.
\end{enumerate}

We have tested the dependence of our results on the choice of fiducial model in Appendix~\ref{app:fid}, where we use the EOS OPT model, which has an enhanced population III contribution to star formation and X-ray production (see~\citealt{Munoz2021b} for details). In the OPT model we change the following parameters: $\fstarII = - 1.25, \fescII = -1.2, \fstarIII = -1.75, \fescIII = -2.35$ and $\nuX=200$ eV. We find our main results are unchanged: the product of $\fstarII \times \fescII$ is still the first principal component, the first four principal components are dominated by population II and X-ray parameters, and PC10 is dominated by $A_\mathrm{LW}$.

\subsection{Inference using the Fisher matrix}
\label{sec:res_inference}

So far we have focused on the ability of the Fisher-matrix formalism to forecast errors with mock data. This can potentially save the running of computationally expensive MCMCs for enlarged parameter spaces (for instance if new parameters such as ETHOS are included,~\citealt{Munoz2020}; Verwohlt et al. in prep.), and re-use the same Fisher matrix as above. However, the same formalism could be used for parameter inference from real data. We now explain the method for inference and its limitations.

The Fisher formalism assumes that the likelihood is Gaussian within the parameter range under consideration. For that, a requirement is that the observable ($\Delta^2_{21}$) changes linearly with the parameters, i.e. $\Delta^2_{21}(k,z) \approx \sum_i \theta_i f_i(k, z)$, where $f_i(k, z)$ are not necessarily linear. Under that assumption, we can infer shifts $\delta \theta_i$ on parameters of a given data set, when compared to our fiducial \citep[e.g.,][]{Munoz2016}. A major caveat is that this assumption of linearity is likely to break down when the observed parameters are a sufficient distance from the fiducial parameters.

We start by writing the observed 21-cm power spectrum as a linear function of the fiducial model power spectrum and the parameter shifts:
\begin{equation} \label{eqn:inference_ps}
    \Delta^2_{21,\mathrm{obs}}(k,z; \{\theta^\prime\}) = \Delta^2_{21,\rm fid} + \sum_i\dfrac{\partial \Delta^2_{21,\rm fid}}{\partial\theta_i} (\theta^\prime_i  - \theta_{i, \rm fid}),
\end{equation}
for a set of parameters $\{\theta^\prime\}$ which differ from the fiducial parameters, $\{\theta_{\rm fid.}\}$.
This formula will be exact only for small parameter differences $(\theta^\prime_i  - \theta_{i, \rm fid} \ll 1)$, but can allow us to quickly test the deviation between the parameters of an observation and those of our fiducial.

Given the difference $\Delta^2_{21,\rm diff} = \Delta^2_{21,\mathrm{obs}}(\theta^\prime) - \Delta^2_{21,\rm fid.}$ between the observed 21-cm PS with parameters $\theta^\prime$, and that of the fiducial, we define the difference vector as
\begin{equation} \label{eqn:inference_Dj}
    D_j =\sum_{i_k,i_z}  \Delta^2_{21,\rm diff}(k,z) \dfrac{\partial \Delta^2_{21,\rm fid}(k,z)}{\partial{\theta_j}} \dfrac{1}{\sigma_{\Delta^2}^2(k,z)}
\end{equation}
akin to Equation~\ref{eqn:Fij_PS} (which defined the Fisher matrix), but with one of the derivatives swapped for $\Delta^2_{21,\rm diff}$.

Then, taking into account the covariance matrix $\mathcal C=\mathcal F^{-1}$ between parameters, we can find the shifts between the fiducial parameters and the observed ones as
\begin{equation} \label{eqn:inference_params}
    \delta \theta_i = \theta^\prime_i  - \theta_{i, \rm fid} = \sum_j \mathcal C_{ij} D_j.
\end{equation}
We test this formalism by comparing two different choices of mock observations against our baseline (EOS21) fiducial one.
\begin{enumerate}
    \item In the first we only slightly modified the parameters, randomly sampling the parameters from within 1$\sigma$ of our expected errors for the EOS fiducial with moderate noise (i.e. they fall within the blue ellipses in Figure~\ref{fig:corner_EOS21_noise}). In this case we expect the likelihood to be roughly Gaussian, and the shifts to be consistent with the inputs.
    \item In the second we aim to infer the OPT EOS model, which has an enhanced population III contribution to star formation and X-ray production (see~\citealt{Munoz2021b} for details). In the OPT model we change the following parameters: $\fstarII = - 1.25, \fescII = -1.2, \fstarIII = -1.75, \fescIII = -2.35$ and $\nuX=200$ eV. These are significantly different from our EOS21 fiducial (see Table~\ref{tab:fiducial_params}), so they will test how far from the fiducial the parameters can be before the formalism breaks down.
\end{enumerate}
In both cases we generate the new mock observed 21-cm power spectra, $\Delta^2_{21,\mathrm{obs}}(\theta^\prime)$, using identical settings as for our fiducial EOS run, changing only the astrophysical parameters.

A subtlety in this analysis is that, as well as the Poisson noise in the power spectrum -- due to measuring the power spectrum from a finite number of $k$-space modes in the simulation boxes -- both our fiducial and ``observed'' mocks have sample (cosmic) variance as they are each independent realizations of the initial conditions. Both the Poisson noise and cosmic variance depend on the volume of the simulations (see e.g., \citealt{Iliev2014,Kaur2020} for the power spectrum and \citealt{Munoz:2020itp} for the global signal). Thus the total noise on the power spectrum is the Poisson, cosmic variance and instrumental noise terms added in quadrature: $\sigma_{\rm tot}^2(k,z) = \sigma_{\rm poiss.}^2(k,z) + \sigma_\mathrm{c.v.}^2(k,z) + \sigma_{\rm instr.}^2(k,z)$.
As such, using the Fisher matrix as defined above undercounts the error bars on our mock observation, as we have assumed the mock has the cosmic variance from a HERA-like cosmic observation \citep[approximately 150\,Gpc$^3$,][]{deBoer2017}, rather than the much smaller volume of our mock simulation (0.064\,Gpc$^3$, see Section~\ref{sec:method_21cmsense}).

To circumvent the issue of cosmic variance, we first run the two cases with the same initial conditions as our EOS fiducial. Using simulations with the same initial conditions as the fiducial we find we can recover the `observed' parameters within $2\sigma$ (7/10 parameters recovered within $1\sigma$) for case  (i), but the inferred parameters are not well-recovered for case (ii), with only 1/10 parameters recovered within $1\sigma$. The biggest offset between the inferred and true values are for parameters where the shift $\delta \theta_i$ is greater than the $1\sigma$ errors on the fiducial value. We surmise that the inference approach outlined above only works well when the assumption that the power spectrum varies linearly with the parameters holds -- i.e. when the likelihood is most like a Gaussian, and the `true` parameters are within $1\sigma$ of the peak likelihood for our specific fiducial. 

When varying the initial conditions we find a worse recovery of the parameter shifts for both cases. This is ameliorated when increasing the instrumental error bars (using our `pessimistic' noise case), as that increases the variance of each $k$ number, which partially accounts for the undercounted cosmic variance. Given that we cannot ``match'' the initial conditions of a real power spectrum detection in the sky, we conclude that using the Fisher matrix for inference is challenging in our formalism, even if the likelihood is approximately Gaussian. Thus a dedicated variance analysis may be required but is beyond the scope of this paper. We note that as our simulation volumes are converged for the power spectrum on the scales we consider \citep[see Section~\ref{sec:method_21cmsense} and e.g.,][]{Kaur2020}, the cosmic variance noise should be subdominant to other sources of noise. Thus the poor performance of the Fisher matrix interference is likely dominated by the non-linearity of the power spectrum as a function of the parameters.

\section{Discussion} 
\label{sec:disc}

\subsection{Computational efficiency} \label{sec:disc_benchmarking}

The key advantage of using Fisher matrices over MCMC is the reduced computational time, due to requiring far fewer individual simulation runs. For the Fisher-matrix calculation we require $2N+1$ simulations, where $N$ is the number of free parameters, whereas for an MCMC with {\tt 21CMMC} of order $\sim10^5$ simulations are required to reach convergence \citep[][]{Greig2017b}. 

For comparison, the MCMC run by \citet{Park2019} required 70,000 individual simulations, including burn-in steps, whereas our Fisher-matrix calculation using the same fiducial parameters required only 17 individual simulations. While the simulation generation can be parallelised, we still expect the Fisher-matrix approach can greatly reduce the computation time compared to an MCMC. The Fisher matrix also has the advantage that additional parameters can be varied by just running a couple of new simulations, rather than having to re-run an MCMC to explore the joint posterior. While an MCMC should be the gold standard for analysis of observations \citep[e.g.,][]{Ghara2020,Ghara2021,Greig2021a,Greig2021b,HERA2021b,Hera2022}, we have demonstrated that a Fisher-matrix analysis can accurately recover the same parameter degeneracies and uncertainties. 

This means a Fisher-matrix approach could be useful for testing and prototyping analyses to forecast parameter uncertainties. Potential applications could be: adding new parameters -- for example cosmological and dark-matter parameters, which will be the topic of a future work (Verwohlt et al. in prep); or testing the impact of instrument/observing designs (described more below). 

The limitations of the Fisher-matrix approach are that, as described in Section~\ref{sec:res_inference}, it is difficult to use the Fisher-matrix approach for actual inference, thus MCMC or nested sampling will still be required to map full posterior when we have 21-cm observations. But, for forecasting, the Fisher-matrix approach is extremely efficient for estimating parameter uncertainties assuming a fiducial model (e.g. using the maximum {\it a posteriori} model from current, non-21-cm observations \citep[e.g.,][]{Park2019,Qin2021b}. Additionally, parameter estimates from Fisher matrices are always optimistic, as the Fisher matrix will always produce the lower bound on the possible observed covariance matrix due to the assumption of a multi-variate Gaussian posterior (see Section~\ref{sec:method_fisher}), so it is possible that some of the uncertainties will be underestimated.

\subsection{Importance of understanding observational noise} \label{sec:disc_noise}

As demonstrated in Section~\ref{sec:res_full}, assumptions about the noise level of the 21-cm observations can have order of magnitude effects on the inferred parameter constraints. Figure~\ref{fig:frac_error} shows the fractional error in each parameter under the assumption of the moderate and pessimistic foreground noise models, with and without the inclusion of the UV LF prior on \astarII.

An advantage of the Fisher-matrix approach is that it is very efficient to rerun the forecast analysis under different assumptions of the foreground noise model, whereas using an MCMC this would require re-calculating the likelihood for every simulation. As a good understanding of the noise is crucial to measuring astrophysics and cosmology from 21-cm observations, our approach provides an efficient way to assess the impact of assumptions in instrumental noise and foreground models.

\section{Conclusions} 
\label{sec:conc}

We have created a Fisher-matrix wrapper for the public code {\tt 21cmFAST}, {\tt 21cmfish}, enabling rapid parameter forecasts from the cosmic 21-cm signal. Our conclusions are as follows:
\begin{itemize}
    \item We verify that our Fisher-matrix analysis recovers the same parameter degeneracies and produces comparable uncertainties as the 21cm-only MCMC of \citet{Park2019}, requiring only $\sim0.03\%$ of the individual simulations. This means that, under the assumption of a multi-variate Gaussian posterior, a Fisher matrix can be used to rapidly explore parameter constraints with 21-cm observations \citep[see also e.g.,][]{Pober2014,Liu2016,Shimabukuro2017,Munoz2020,Jones2021}, significantly reducing the requirement for expensive and energy intensive computation when prototyping analyses \citep[and does not require training, c.f. emulator approaches e.g.,][]{Kern2017}.
    \item Using our Fisher-matrix approach, we perform the first joint analysis of population II and population III galaxy parameters as well as those characterizing feedback from radiative backgrounds. We find that under the assumption of a `moderate' foreground noise floor (using the `moderate' foregrounds model in {\tt 21cmsense}, with a super-horizon buffer $a=0.1\,h \Mpcinv$ and expected HERA system temperature, \citealt{deBoer2017}), both population II and population III parameters can be constrained by future HERA data to $\simlt10\%$ precision, due to the relative importance of the parameters as a function of redshift and the wide frequency coverage of HERA.
    \item Adding priors on parameters from independent measurements of the UV luminosity function from Hubble Space Telescope data to 21-cm observations improves constraints on population II parameters by approximately a factor of three in the case of pessimistic foreground noise, but does not significantly improve estimates in the case of moderate foreground noise as those constraints are already comparable to the UV LF information.
    \item A principal component analysis demonstrates that at least 7 combinations of the parameters could be well-measured by HERA. The first four principal components are dominated by $\fescII\fstarII$ and combinations of population II and X-ray emission properties, as these are the most important features in determining the strength of the 21-cm signal at $z\simlt10$ where the signal will be best measured. Our analysis shows the ionizing photon escape fraction and stellar-to-halo mass fraction are highly degenerate using 21-cm observations alone \citep[see also,][]{Park2019}.
    \item We attempt to use our Fisher-matrix approach to perform an inference of simulated data and find that it is only possible to reliably recover parameters that were within $1\sigma$ of the fiducial simulation and when using the same initial conditions for the fiducial and mock data simulations. Thus, we caution that while the Fisher matrix is useful for forecasting, within our formalism it is not a good choice for inference: Bayesian inference techniques such as MCMC or Nested sampling are more appropriate for mapping out the full posterior for unknown parameters, once detections are available.
\end{itemize}
Our results show that Fisher-matrix analyses provide a fast and realistic way to estimate parameter uncertainties from future 21-cm observations, under the assumption the posterior is a multi-variate Gaussian and also assuming the noise is separable from the cosmic signal.
This greatly reduces the requirement for expensive MCMC runs when prototyping new analyses or adding additional parameters. This framework enables us to more easily explore the effects of foreground noise and the degeneracies between astrophysical and cosmological, e.g. dark-matter, parameters, which will be the topic of a future work.


\section*{Acknowledgements}
CAM acknowledges support by the VILLUM FONDEN under grant 37459, the Carlsberg Foundation under grant CF22-1322, and from NASA Headquarters through the NASA Hubble Fellowship grant HST-HF2-51413.001-A awarded by the Space Telescope Science Institute, which is operated by the Association of Universities for Research in Astronomy, Inc., for NASA, under contract NAS5-26555. The Cosmic Dawn Center (DAWN) is funded by the Danish National Research Foundation under grant DNRF140.
JBM acknowledges support by a Clay Fellowship at the Smithsonian Astrophysical Observatory.
Parts of this research were supported by the Australian Research Council Centre of Excellence for All Sky Astrophysics in 3 Dimensions (ASTRO 3D), through project number CE170100013. JP was supported by a KIAS individual Grant (PG078701) at Korea Institute for Advanced Study.

\section*{Data availability}
Our Fisher analysis package for {\tt 21cmFAST} is available in a public python package {\tt 21cmfish}: \url{https://github.com/charlottenosam/21cmfish}. The code and accompanying data to reproduce plots in this paper are included in the package.

\section*{Software}
{\tt 21cmFAST} \citep{Mesinger2011,Park2019,Qin2020,Murray2020,Munoz2021b}, {\tt 21cmSense} \citep{Pober2014}, {\tt corner} \citep{corner}, {\tt IPython} \citep{Perez2007a}, {\tt matplotlib} \citep{Hunter2007a}, {\tt NumPy} \citep{VanderWalt2011a}, {\tt SciPy} \citep{Oliphant2007a}, {\tt Astropy} \citep{Robitaille2013}.


\appendix

\section{Dependence on choice of fiducial model} \label{app:fid}

\begin{figure}
\includegraphics[width=\columnwidth]{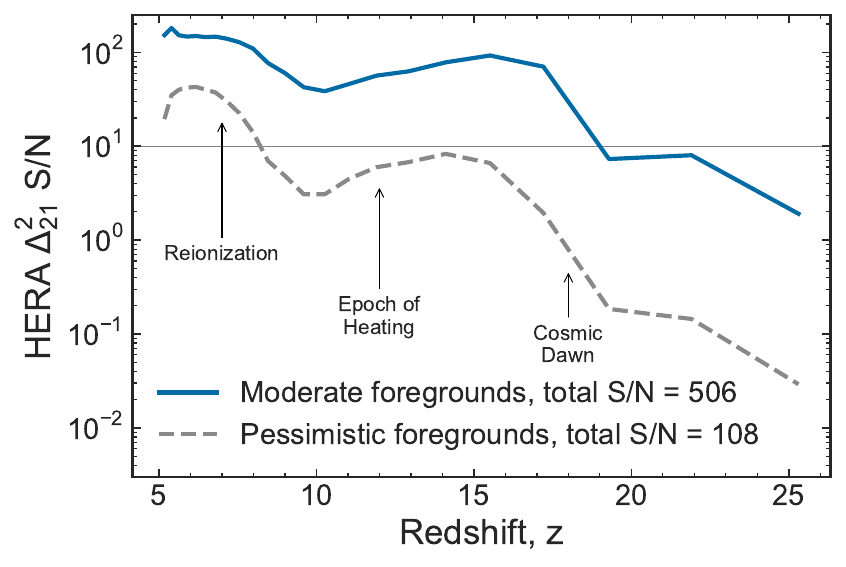}
\caption{Same as Figure~\ref{fig:EOS21_SNR} but for the EOS OPT fiducial model
\label{fig:EOS21_SNR_OPT}
}
\end{figure}

\begin{figure*}
\includegraphics[width=\textwidth]{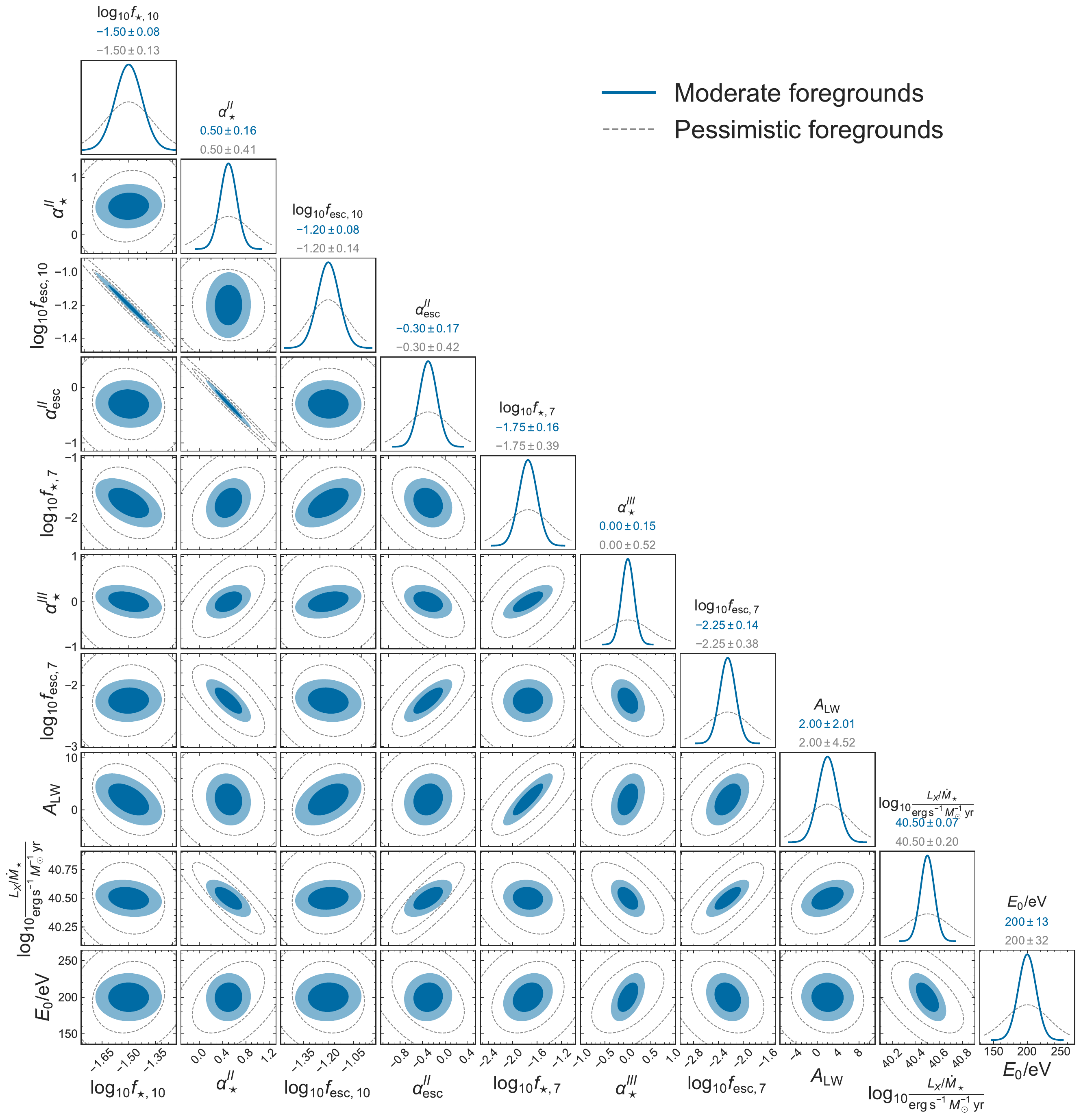}
\caption{Same as Figure~\ref{fig:corner_EOS21_noise} but for the EOS OPT fiducial model
\label{fig:corner_EOS21_OPT}
}
\end{figure*}

\begin{figure}
\includegraphics[width=\columnwidth]{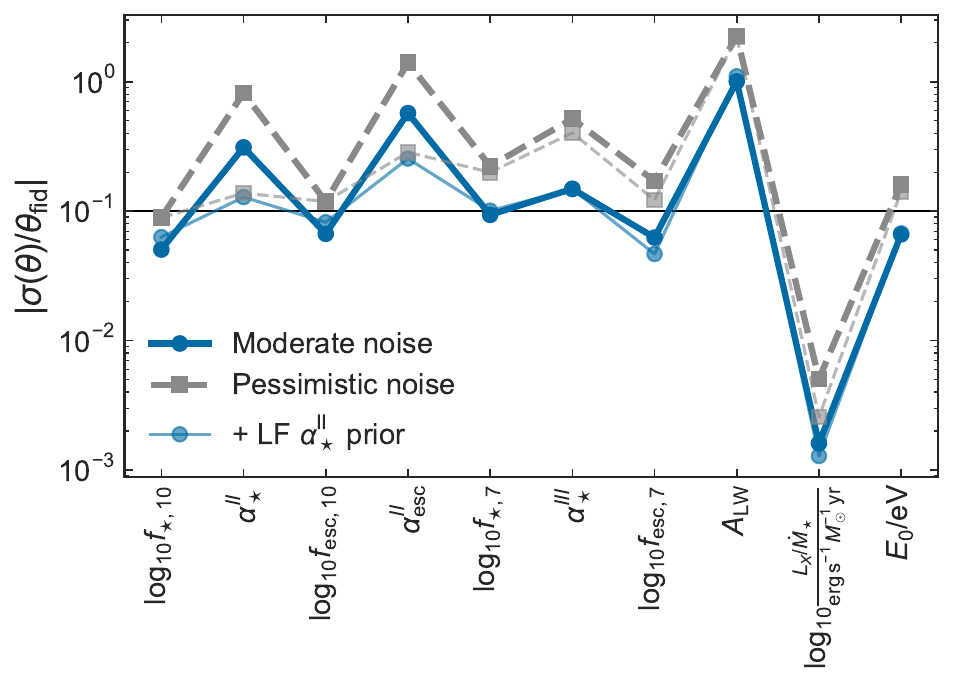}
\caption{Same as Figure~\ref{fig:frac_error} but for the EOS OPT fiducial model
\label{fig:frac_error_OPT}
}
\end{figure}

\begin{figure}
\includegraphics[width=\columnwidth]{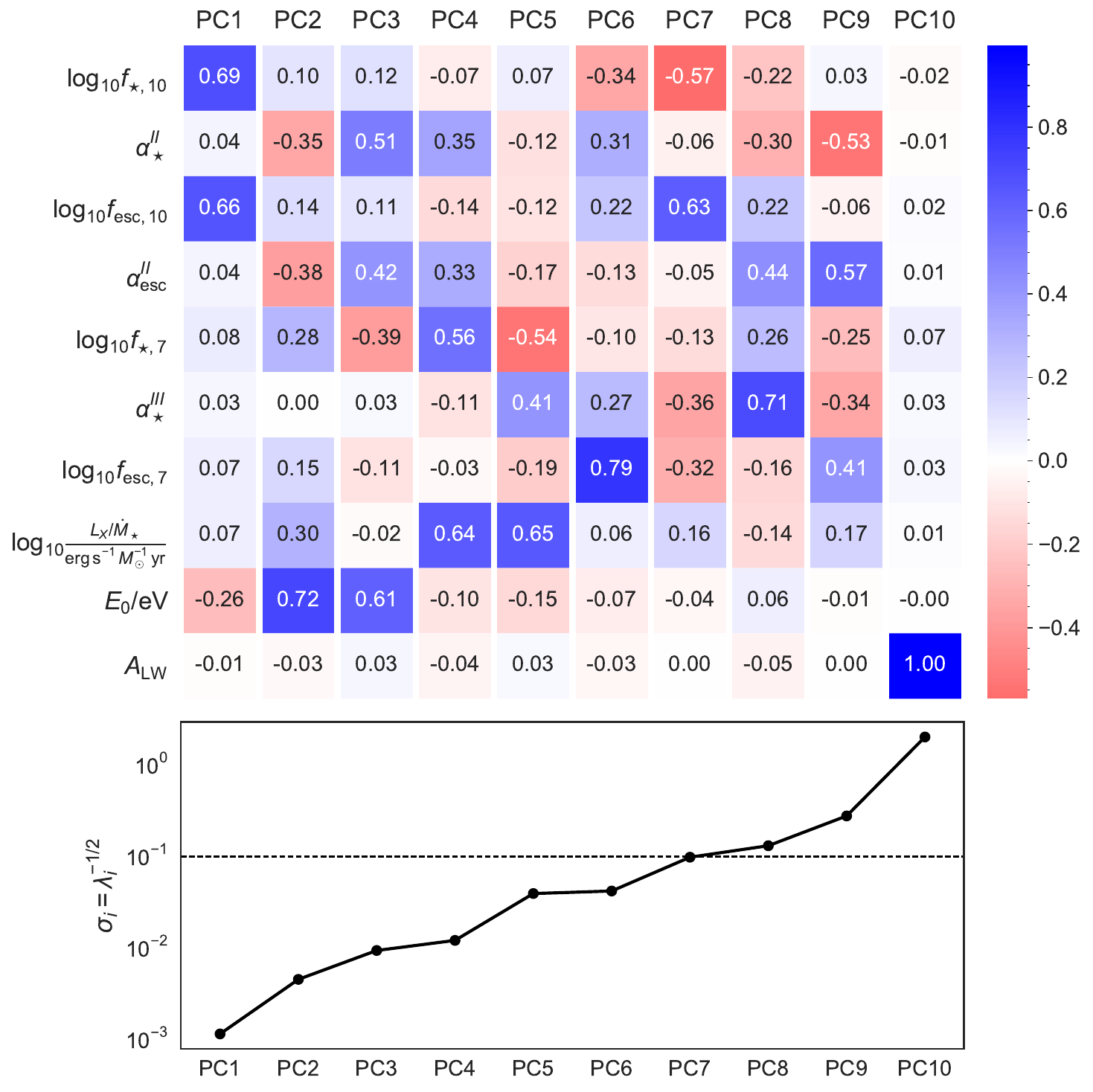}
\caption{Same as Figure~\ref{fig:PC_sigma} but for the EOS OPT fiducial model
\label{fig:PC_sigma_OPT}
}
\end{figure}

We do not know the model that best describes the data \textit{a priori}. In the above we have used a fiducial model which uses constraints from existing observations (e.g. the cosmic SFR density and the reionization timeline) to inform the astrophysical parameters of our model. The exact precision with which we can measure astrophysical parameters (Section~\ref{sec:res_full}) and the combination of parameters which can be best-constrained (Section~\ref{sec:res_full_PCA}) will depend on the fiducial model, as the relative strength of the population II and III galaxy components and of X-ray heating will all change the 21-cm signal. However, because these effects change the signal in different ways (as demonstrated by our PCA analysis in Section~\ref{sec:res_full_PCA}) we expect our results will not be strongly affected qualitatively by the choice of fiducial model.

To assess this in more detail we carry out the same analysis as in Sections~\ref{sec:res_full} and \ref{sec:res_full_PCA} for a second fiducial model, the EOS OPT model \citep[also used in Section~\ref{sec:res_inference}, see][]{Munoz2021b} which has an enhanced contribution of population III galaxies to star formation and enhanced X-ray emission compared to our original model. In the OPT model we change the following parameters: $\fstarII = - 1.25, \fescII = -1.2, \fstarIII = -1.75, \fescIII = -2.35$ and $\nuX=200$ eV. We find our results are qualitatively unchanged.

We show the signal-to-noise ratio of this 21-cm power spectrum in Figure~\ref{fig:EOS21_SNR_OPT}, where we see an increased S/N, particularly during the Epoch of Heating due to the higher X-ray heating, compared to our EOS fiducial model (Figure~\ref{fig:EOS21_SNR}). The parameter constraint forecast is plotted in Figure~\ref{fig:corner_EOS21_OPT} and the fractional error on each parameter in Figure~\ref{fig:frac_error_OPT}. Because the S/N ratio is higher for this model, we find less of a difference between the moderate and pessimistic foregrounds. Because of the increased population III galaxy contribution to the cosmic SFR density, we find population III galaxy parameters could be slightly better constrained than in our original fiducial, while population II galaxy parameters would be slightly less well constrained because there are degeneracies between parameters governing the two galaxy populations

Finally, we show the principal components of the 21-cm power spectrum in Figure~\ref{fig:PC_sigma_OPT}. The principal components are very similar to those of the EOS model (Figure~\ref{fig:PC_sigma}). The product of $\fstarII \times \fescII$ is still the first principal component, the first four principal components are dominated by population II and X-ray parameters (with X-ray parameters contributing slightly more to the PCs compared to in the EOS model), and $A_\mathrm{LW}$ is still the last principal component.

\section{Principal component analysis in redshift windows} \label{app:PCA}

Our principal component analysis in Section~\ref{sec:res_full_PCA} demonstrates the combinations of parameters that will be best-constrained by the full HERA dataset. However, due to the strong frequency and therefore redshift dependence of the noise (see Figure~\ref{fig:EOS21_SNR}), where the background is higher at lower frequencies \citep[e.g.,][]{deBoer2017}, the principal components are dominated by the parameters most important for the Epoch of Reionization -- at $z\simlt10$ where the noise is lowest.

To provide more insight into the redshift dependence of the principal components we do a PCA in three redshift bins, corresponding to: the Epoch of Reionization ($z\sim5-10$), where the global 21-cm signal peaks and declines to zero); the Epoch of Heating ($z\sim10-15$), the signal goes from absorption to emission as the IGM is heated (likely by X-ray emission from the first galaxies); and early Cosmic Dawn ($z\sim15-30$), where the global signal starts to drop due to feedback from the first stars, until it reaches the trough before heating dominates.

\begin{figure*}
\includegraphics[width=0.49\textwidth]{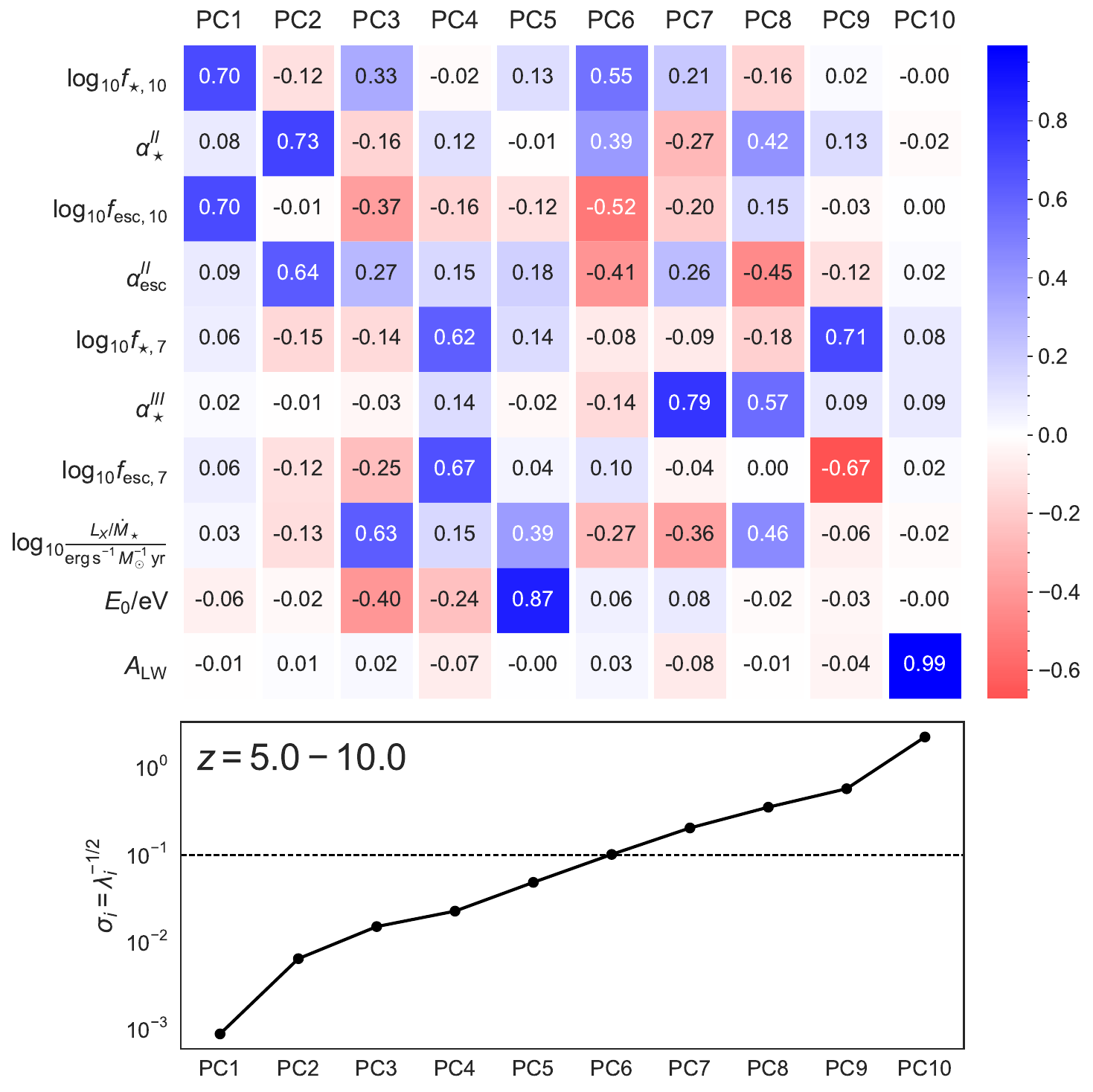}
\includegraphics[width=0.49\textwidth]{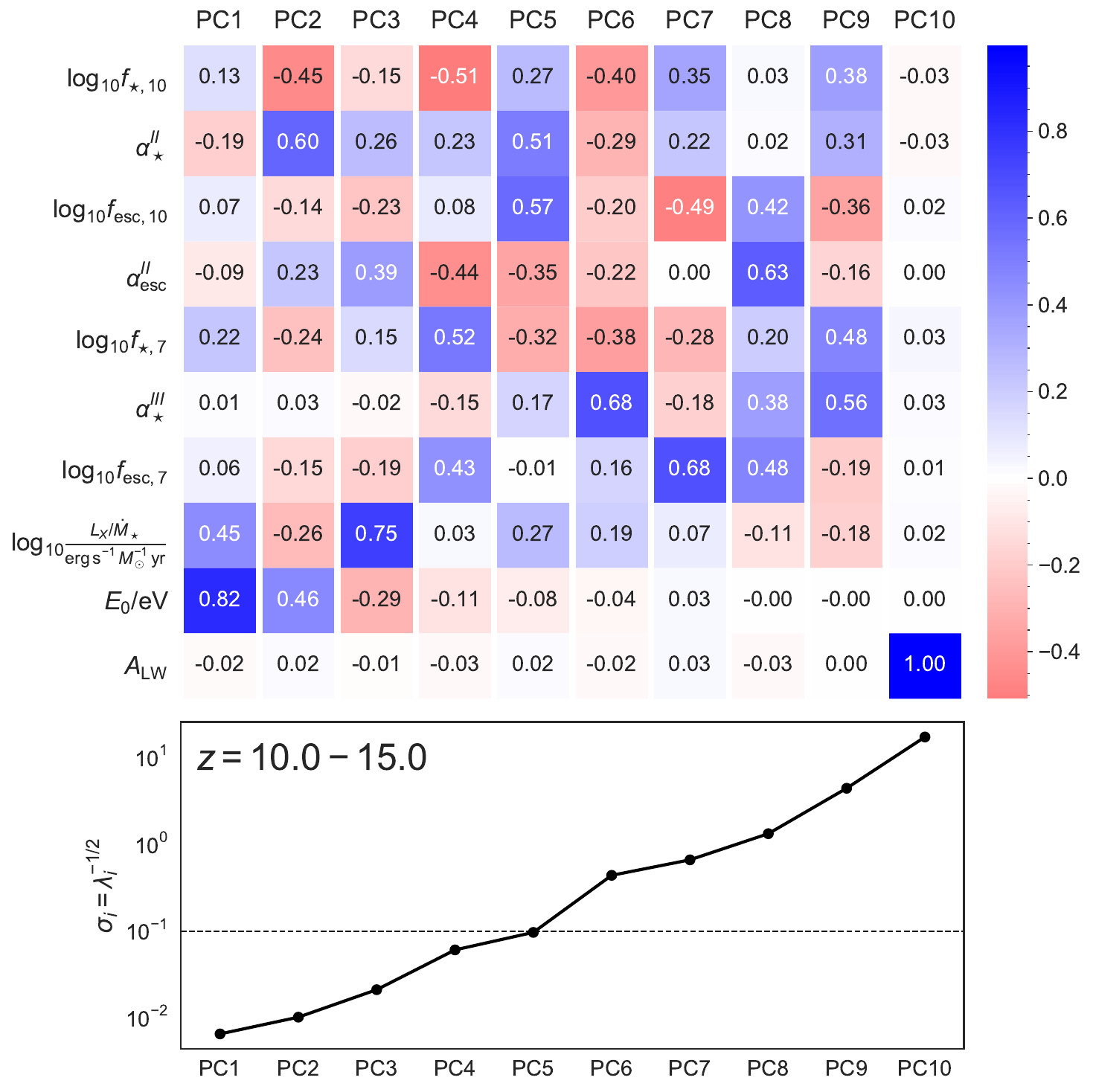}
\includegraphics[width=0.49\textwidth]{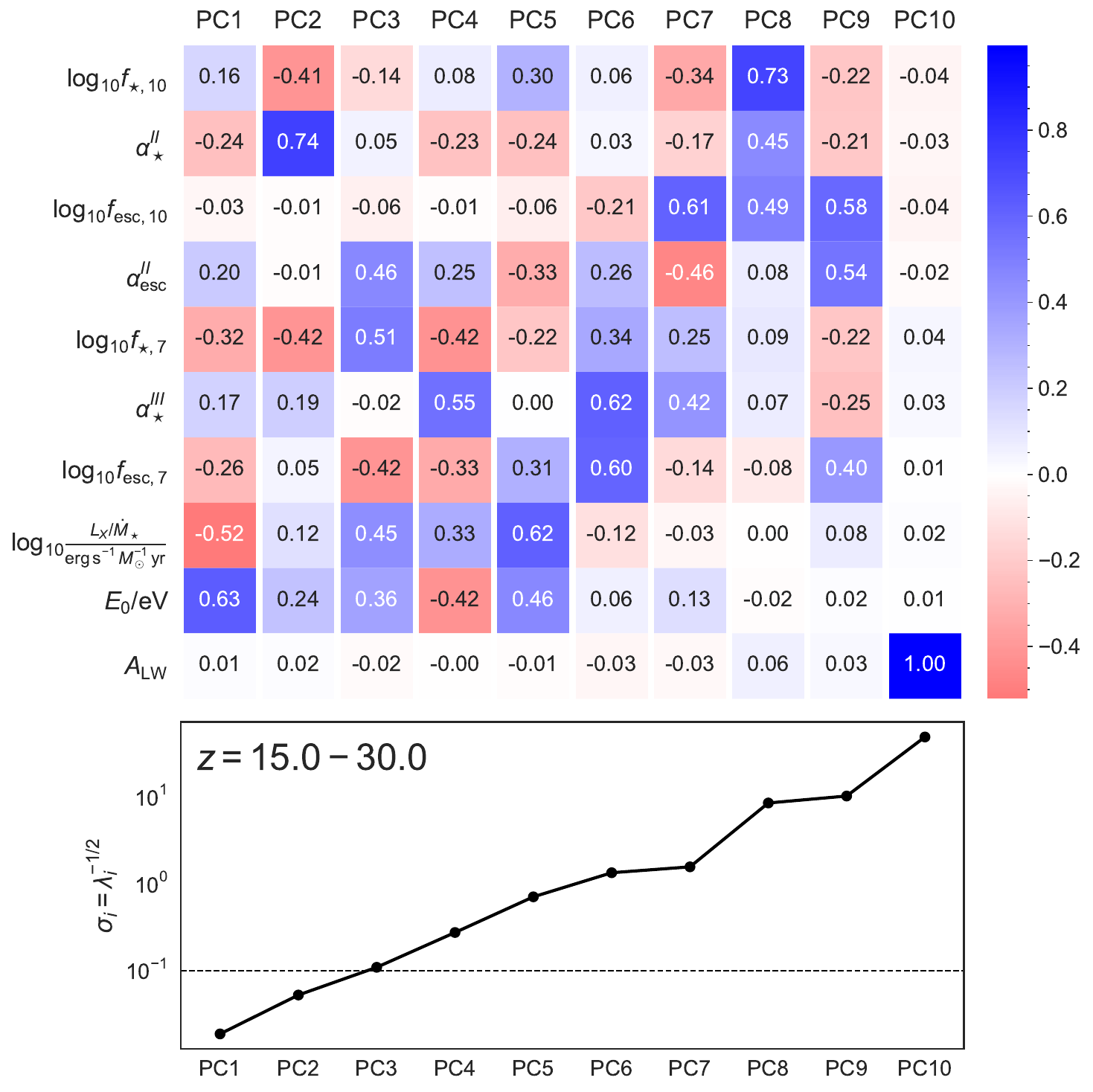}
\caption{(Upper Left) Eigenvectors associated with each principal component, in order of decreasing importance/increasing uncertainty, in the Epoch of Reionization redshift window $z\sim5-10$.
(Upper Right) Eigenvectors associated with each principal component, in order of decreasing importance/increasing uncertainty, in the Epoch of Heating redshift window $z\sim10-15$.
(Lower) Eigenvectors associated with each principal component, in order of decreasing importance/increasing uncertainty, in the early Cosmic Dawn redshift window $z\sim15-30$.
\label{fig:PC_sigma_z}
}
\end{figure*}

We plot the principal components from these three redshift bins in Figure~\ref{fig:PC_sigma_z}. These plots demonstrate which combinations of parameters the observed redshift windows are most sensitive to. For the Reionization redshift window, we see the first principal component is very similar to the one obtain for the whole redshift range: $\fstarII \times \fescII$. The following principal components are dominated by population II and X-ray parameters, as expected during the Reionization epoch. In the Epoch of Heating the primary principal components are dominated by X-ray parameters, as expected, as in the model, X-rays are the dominant driver of IGM heating. For the Cosmic Dawn redshift window, the first principal components are dominated by X-ray parameters and a combination of population III parameters.

\bibliographystyle{mnras}
\bibliography{library} 



\bsp	
\label{lastpage}
\end{document}

%% file: macros.tex
\newcommand{\simgt}{\,\rlap{\lower 3.5 pt \hbox{$\mathchar \sim$}} \raise 1pt \hbox {$>$}\,}
\newcommand{\simlt}{\,\rlap{\lower 3.5 pt \hbox{$\mathchar \sim$}} \raise 1pt \hbox {$<$}\,}

\newcommand{\BE}{\begin{equation}}
\newcommand{\EE}{\end{equation}}
\newcommand{\BEA}{\begin{eqnarray}}
\newcommand{\EEA}{\end{eqnarray}}

\newcommand{\DV}{\ifmmode{\Delta v}\else $\Delta v$\xspace\fi}

\newcommand{\HI}{\ifmmode{\textsc{hi}}\else H\textsc{i}\fi\xspace}
\newcommand{\HII}{\ifmmode{\textsc{hii}}\else H\textsc{ii}\fi\xspace}

\newcommand{\Msun}{\ifmmode{M_\odot}\else $M_\odot$\xspace\fi}
\newcommand{\MUV}{\ifmmode{M_\textsc{uv}}\else $M_\textsc{uv}$\xspace\fi}
\newcommand{\fesc}{\ifmmode{f_\textrm{esc}}\else $f_\textrm{esc}$\xspace\fi}
\newcommand{\lya}{\ifmmode{\mathrm{Ly}\alpha}\else Ly$\alpha$\xspace\fi}

\newcommand{\nh}[1][]{\ifmmode{\overline{n}_\textsc{h}^{#1}}\else $\overline{n}_\textsc{h}$\xspace\fi}

\newcommand{\xHI}{\ifmmode{x_\HI}\else $x_\HI$\xspace\fi}
\newcommand{\xHImean}{\ifmmode{\overline{x}_\HI}\else $\overline{x}_\HI$\xspace\fi}
\newcommand{\xHIImean}{\ifmmode{\overline{x}_\HII}\else $\overline{x}_\HII$\xspace\fi}
\newcommand{\trec}{\ifmmode{t_\textrm{rec}}\else $t_\textrm{rec}$\xspace\fi}
\newcommand{\clump}[1][]{\ifmmode{C_\HII^{#1}}\else $C_\HII$\xspace\fi}

\newcommand{\Nion}{\ifmmode{\dot{N}_{\mathrm{ion}}}\else $\dot{N}_\mathrm{ion}$\xspace\fi}
\newcommand{\Rion}[1][]{\ifmmode{R_\mathrm{ion}^{#1}} \else $R_\mathrm{ion}$\xspace\fi}

\newcommand{\kp}{\ifmmode{k_\textrm{peak}}\else $k_\textrm{peak}$\xspace\fi}
\newcommand{\hp}{\ifmmode{h_\textrm{peak}}\else $h_\textrm{peak}$\xspace\fi}
\newcommand{\hMpc}{\ifmmode{\,h^{-1}\textrm{Mpc}}\else \,$h^{-1}$Mpc\xspace\fi}

\newcommand{\Tb}{\ifmmode{T_{21}}\else $T_{21}$\xspace\fi}
\newcommand{\aesc}{\ifmmode{\alpha_\mathrm{esc}}\else $\alpha_\mathrm{esc}$\xspace\fi}
\newcommand{\fescII}{\ifmmode{f_\mathrm{esc,10}^\textsc{ii}}\else $f_\mathrm{esc,10}^\textsc{ii}$\xspace\fi}
\newcommand{\fescIII}{\ifmmode{f_\mathrm{esc,7}^\textsc{ii}}\else $f_\mathrm{esc,7}^\textsc{iii}$\xspace\fi}
\newcommand{\astarII}{\ifmmode{\alpha_\star^\textsc{ii}}\else $\alpha_\star^\textsc{ii}$\xspace\fi}
\newcommand{\astarIII}{\ifmmode{\alpha_\star^\textsc{iii}}\else $\alpha_\star^\textsc{iii}$\xspace\fi}
\newcommand{\fstarII}{\ifmmode{f_{\star,10}^\textsc{ii}}\else $f_{\star,10}^\textsc{ii}$\xspace\fi}
\newcommand{\fstarIII}{\ifmmode{f_{\star,7}^\textsc{iii}}\else $f_{\star,7}^\textsc{iii}$\xspace\fi}
\newcommand{\tstar}{\ifmmode{t_\star}\else $t_\star$\xspace\fi}
\newcommand{\Mturn}{\ifmmode{M_\mathrm{turn}}\else $M_\mathrm{turn}$\xspace\fi}
\newcommand{\LX}{\ifmmode{L_X/{\dot{M}_\star}}\else $L_X/{\dot{M}_\star}$\xspace\fi}
\newcommand{\nuX}{\ifmmode{E_0}\else $E_0$\xspace\fi}
\newcommand{\AVCB}{\ifmmode{A_\mathrm{VCB}}\else $A_\mathrm{VCB}$\xspace\fi}
\newcommand{\ALW}{\ifmmode{A_\mathrm{LW}}\else $A_\mathrm{LW}$\xspace\fi}
\newcommand{\Mpcinv}{\ifmmode{\,\mathrm{Mpc}^{-1}}\else \,Mpc$^{-1}$\xspace\fi} 

\newcommand{\kms}{\,\ifmmode{\mathrm{km}\,\mathrm{s}^{-1}}\else km\,s${}^{-1}$\fi\xspace}
\newcommand{\cm}{\,\ifmmode{\mathrm{cm}}\else cm\fi\xspace}

